\documentclass{article}

\usepackage{lmodern}
\usepackage{geometry}
\usepackage[utf8]{inputenc} % allow utf-8 input
\usepackage[T1]{fontenc}    % use 8-bit T1 fonts

\usepackage{tkz-tab}
\usepackage{float} 
\usepackage{url}            % simple URL typesetting
\usepackage{booktabs}       % professional-quality tables
\usepackage{amsfonts}       % blackboard math symbols
\usepackage{nicefrac}       % compact symbols for 1/2, etc.
\usepackage{microtype}      % microtypography
\usepackage{amsmath}
\usepackage{lipsum}
\usepackage{graphicx}
\usepackage{xcolor}
\definecolor{C0}{HTML}{3182ce}
\definecolor{C1}{HTML}{dd6b20}
\usepackage[breaklinks=true,colorlinks,bookmarks=false,citecolor=C0]{hyperref}

\usepackage{cancel}
\usepackage{mathtools} % for equal def symbols, \coloneqq and \eqqcolon
\usepackage{booktabs} % for professional tables
\usepackage{multirow}

\usepackage{array}

\usepackage{adjustbox}

\usepackage{caption}
\usepackage{subcaption}

\newcounter{appsection}
\def\appendixname{Appendix}
\renewcommand\appendix{\par
  \setcounter{appsection}{0}%
  \setcounter{subsection}{0}%
  \setcounter{equation}{0}
  \gdef\thefigure{\Alph{appsection}.\arabic{figure}}%
  \gdef\thetable{\Alph{appsection}.\arabic{table}}%
    \gdef\thesection{\appendixname~\Alph{appsection}}%
}

\graphicspath{ {./images/} }

\usepackage{natbib}
\counterwithin*{Lem}{section}

\counterwithin*{LemApp}{section}

\counterwithin*{Theo}{section}

\newtheorem{Prop}{Proposition}

\counterwithin*{Prop}{section}

\counterwithin*{Def}{section}

\newtheorem{Cor}{Corollary}

\counterwithin*{Cor}{section}

\definecolor{forestgreen}{rgb}{0.13, 0.55, 0.13}
\definecolor{darkblue}{rgb}{0.0, 0.0, 0.65}
\definecolor{violet2}{rgb}{0.5,0,0.5}
\definecolor{orange2}{rgb}{0.8, 0.1, 0.1}       
\definecolor{red2}{rgb}{1, 0.4, 0}

\definecolor{colorblind1}{rgb}{0, 0.47, 0.73}
\definecolor{colorblind2}{rgb}{0, 0.6, 0.53}       
\definecolor{colorblind3}{rgb}{0.93, 0.47, 0.2}
\definecolor{colorblind4}{rgb}{0.93, 0.2, 0.47}

\newcommand{\cor}[1]{#1}
\newcommand{\corr}[1]{#1}

\newcommand{\dd}{\mathrm{d}}

\title{Reparameterization of extreme value framework \\ for improved Bayesian workflow}

\author{
    Théo Moins \thanks{Univ. Grenoble Alpes, Inria, CNRS, Grenoble INP, LJK, 38000 Grenoble, France.}
    \and Julyan Arbel \footnotemark[1]
    \and Stéphane Girard \footnotemark[1]
    \and Anne Dutfoy \thanks{EDF R\&D dept. Périclès, 91120 Palaiseau, France.}
}

\begin{document}
\maketitle

\begin{abstract}
    
Using Bayesian methods for extreme value analysis offers an alternative to frequentist ones, with several advantages such as easily dealing with parametric uncertainty or studying irregular models.
However, computation\corr{s} can be challenging and the efficiency of algorithms can be altered by \corr{poor parametrization choices}.
\corr{The focus is on} the Poisson process characterization of univariate extremes and outline two key benefits of an orthogonal parameterization.
First, Markov chain Monte Carlo convergence is improved when applied on orthogonal parameters.
This analysis relies on convergence diagnostics computed on several simulations.
Second, orthogonalization also helps deriving Jeffreys and penalized complexity priors, and establishing posterior propriety thereof.
\corr{The proposed framework} is applied to return level estimation of Garonne flow data (France).

\end{abstract}

\section{Introduction}

Studying the long-term behavior of environmental variables is necessary to understand the risks of hazardous meteorological events such as floods, storms, or droughts.
To this end, models from extreme value theory allow \cor{us} to extrapolate data in the distribution \cor{tails}, in order to estimate extreme quantiles that may not have been observed \citep[see][for an introduction]{Coles2001}.
In particular, a key quantity to estimate is  \cor{the} return level $\ell_T$ associated with a given period of $T$ years, the level that is exceeded \cor{on} average once every $T$ years.
Assessing the resistance of facilities to natural disasters such as dams to floods that occur \cor{on} average once every $100$ years or $1\,000$ years is critical for companies \cor{such as} \cor{\textit{\'Electricité de France}} (EDF). 
Moreover, characterizing the uncertainty on the estimation of this return level is also of interest, which encourages the choice of the Bayesian paradigm. 
However, \cor{performing} Bayesian inference requires multiple steps that must be managed by the user, from the choice of the model to the evaluation and validation of computations.
This has been recently formalized by \cite{gelman2020bayesian} in the form of a Bayesian workflow.
After introducing models stemming from extreme value theory in Section~\ref{subsec:bayes_ext_intro}, we briefly review in Section~\ref{subsec:reparam} one particular step of the workflow, reparameterization, and more specifically the choice of an orthogonal parameterization.

\subsection{Extreme-value models}
\label{subsec:bayes_ext_intro}

Three different frameworks exist to model extreme events, leading to different likelihoods: one by block maxima, one by peaks-over-threshold, and one that unifies both \cor{through} a Poisson process characterization.

\paragraph{Block maxima model}
Let $M_n$ be the maximum of $n$ i.i.d random variables with cumulative distribution function (cdf) $F$.
We assume that $F$ belongs to the maximum domain of attraction \cor{of a non-degenerate cdf $G$, meaning} that there exist two sequences $a_n > 0$ and $b_n$  such that 
\cor{$(M_n-b_n)/a_n$ converges in distribution to the cdf $G$.}
The extreme value theorem \citep[\corr{e.g., }][Chapter~1]{de2006extreme} states that $G$ is necessarily a generalized extreme-value (GEV) distribution, with cdf:
\begin{equation}
 G(x)=
  \begin{cases}
   \exp\left(-\left\{1 + \xi
  x\right\}_+
   ^ {-{1}/{\xi}}\right)
   &\quad \text{if $\xi \neq 0$} \,,\\
   \exp(-\exp(-x))
   &\quad \text{if $\xi = 0$},
  \end{cases}
 \label{eq:GEV-CDF}
\end{equation}
where $\{x\}_+ = \max\{0, x\}$. 
Consequently, for a finite value of $n$, one can consider the approximation
$
{\mathbb P}(M_n \leq x) \approx G((x-b_n)/a_n)=:G(x \mid b_n, a_n, \xi),
$
and focus on the estimation of the three parameters of the GEV distribution.
\cor{Here, as the dataset is fixed, the dependence in $n$ for the location and scale parameters will be omitted.}
To obtain a sample of maxima, \cor{one can divide} the dataset into $m$ blocks of size $n/m$ and extract the maximum from each of them.

\paragraph{Peaks-over-threshold model}

Alternatively, one can consider observations that exceed a high threshold $u$.
\cor{Let $X$ be a random variable with cdf $F$.}
Pickands theorem \citep{pickands1975statistical} states that, if $F$ belongs to the maximum domain of attraction of $G$ with ${\mathbb{P}}(M_n \leq x) \approx G(x \mid \mu, \sigma, \xi)$, then the distribution of the exceedances $X-u \mid X>u$ is, as $u$ converges to the upper endpoint of $F$, a generalized Pareto distribution (GPD), with cdf
\begin{equation}
  H(y \mid \Tilde{\sigma}, \xi)=
    \begin{cases}
        1 - \left\{1 + \xi \frac{y}{\Tilde{\sigma}} \right\}_+^{-{1}/{\xi}}
        &\quad \text{if $\xi \neq 0$} \,,\\
        1 - \exp\left(-\frac{y}{\Tilde{\sigma}}\right)
        &\quad \text{if $\xi = 0$},
    \end{cases}
    \label{eq:gpd_distribution}
\end{equation}
where the shape parameter $\xi$ is the same as in~\eqref{eq:GEV-CDF} and the GPD and GEV scales are linked by $\Tilde{\sigma} = \sigma + \xi (u - \mu)$.
%Here, 
To obtain a sample of $n_u$ excesses, the peaks-over-threshold method focusses on the $n_u$ largest values of the dataset.
\corr{It thus requires the estimation of the quantile of order $1-n_u/n$}, which can be seen as the third parameter to estimate, in addition to $\Tilde{\sigma}$ and $\xi$. The most classical choice is to 
estimate this intermediate quantile by the $(n-n_u)$th order statistic.

\paragraph{Poisson process characterization of extremes}
Finally, these two approaches can be generalized by a third one, using a non-homogeneous Poisson process.
We present here an intuitive way for obtaining this model similarly to \citet[Chapter~7]{Coles2001}, and refer to \cite[Chapter~5]{leadbetter2012extremes} for theoretical details.
%more details on point process theory and technical details associated with this construction.
We start by observing that, for large $n$, $F^n(x) \approx G(x \mid \mu, \sigma, \xi)$, for $x$ in the support of $G$ denoted by $\text{supp}(G(\cdot \mid \mu, \sigma, \xi)) = \left\{ x \in \mathbb{R} \text{\quad s.t. } 1+\xi\left(\frac{x-\mu}{\sigma} \right) > 0 \right\}$. 
Hence, considering a large threshold $u \in \text{supp}(G(\cdot \mid \mu, \sigma, \xi))$, a Taylor expansion yields 
\begin{equation*}
    n \log F(u) \simeq -n (1-F(u)) \simeq \log G(u \mid \mu, \sigma, \xi),
\end{equation*}
or, equivalently, 
\begin{equation}
    \mathbb{P}\left(X > u \right) \simeq -\frac{1}{n}  \log G(u \mid \mu, \sigma, \xi).
    \label{eq:excess_prob}
\end{equation}
Equation~(\ref{eq:excess_prob}) can be seen as the probability of $X$ to belong to $I_u := [u, +\infty)$.
In the case of $n$ i.i.d random variables, \corr{one} can deduce that the associated point process $N_n$ is such that 
$N_n(I_u) \sim \mathcal{B}(n,p_n)$ with $p_n$ given by Equation~(\ref{eq:excess_prob}). 
As $n \rightarrow \infty$, the binomial distribution $\mathcal{B}(n,p_n)$ converges to the Poisson distribution $\mathcal{P}(\Lambda(I_u))$, with $\Lambda(I_u) = -\log G(u \mid \mu, \sigma, \xi)$.
This property being valid for all $I_u$ together with the independence property on non-overlapping sets imply that $N_n$ converges to a non-homogeneous Poisson process, with intensity measure $\Lambda(I_u)$:
$N_n \xrightarrow[]{d} N$, with $N(I_u) \sim \mathcal{P}(\Lambda(I_u))$.
This model generalizes the block maxima one since
\begin{equation*}
    \mathbb{P}(M_n < x) = \mathbb{P}(N_n(I_x) = 0) \to
    \mathbb{P}(N(I_x) = 0) = \exp (-\Lambda(I_x)) = G(x \mid \mu, \sigma, \xi) ,
\end{equation*}
as $n\to\infty$.
However, an estimation of the parameters $(\mu, \sigma, \xi)$ with this model is related to the overall maximum  $M_n$ of the dataset, and it is frequent to study maxima of $m$ smaller blocks $M_{n/m}$, where $m$ is typically the number of years in the observations and so $M_{n/m}$ corresponds to annual maxima. 
To do so, the intensity measure is multiplied by $m$, which modifies the parameterization and in particular \cor{the value of $\mu$ and $\sigma$}:
\cite{Wadsworth2010} shows that, if \cor{$(\mu_{k_i}, \sigma_{k_i}, \xi)$ ($i \in\{ 1, 2$\})}, are parameters for $k_i$ GEV observations, then 
\begin{equation}
 \mu_{k_2} = \mu_{k_1} - \frac{\sigma_{k_1}}{\xi}
  \left( 1 - \left(\frac{k_2}{k_1}\right) ^ {-\xi} \right), \quad
 \sigma_{k_2} = \sigma_{k_1} \left(\frac{k_2}{k_1}\right) ^ {-\xi}.  \label{eq:mu_m}
 %\label{eq:sigma_m}
\end{equation}
The threshold excess model can also be derived from the point process representation, since $\mathbb{P}(X > y + u \mid X > u) \simeq  1-H(y \mid \Tilde{\sigma}, \xi)$, 
with $\Tilde{\sigma} = \sigma + \xi (u - \mu)$. 
Moreover, in contrast to the peaks-over-threshold model where an intermediate quantile needs to be estimated, the Poisson model \cor{directly} includes  a third location parameter~$\mu$.

In the following, we will focus \cor{mainly} on this latter model, and treat the peaks-over-threshold \cor{method} as a special case in Section~\ref{subsec:simulations}. 

\paragraph{Bayesian inference}

Using the Bayesian paradigm in extreme value models \cor{is advantageous} in comparison to the frequentist approach, see \cite{ColesPowell1996} for a general review, and \cite{Stephenson2016} or \cite{Bousquet2021} for more recent overviews.
For the Poisson process characterization of extremes, Bayesian inference consists in fixing a scaling factor $m$ and a threshold $u$ to get \corr{a number of $n_u\geq 1$ observations exceeding $u$ denoted by} $\boldsymbol{x} = (x_1, \ldots, x_{n_u})$.
The likelihood of these observations can be written as
    \begin{equation}
    % \textstyle
        L(\boldsymbol{x}, n_u \mid \mu, \sigma, \xi)
        = \mathrm{e}^{-m \left(1 + \xi \left(\frac{u - \mu}{\sigma}\right) \right) ^{-1/\xi}}
        \sigma^{-n_u} \prod_{i = 1}^{n_u} \left(1 + \xi \left(\frac{x_i - \mu}{\sigma}\right) \right) ^{-1-1/\xi}.
        \label{eq:likelihoodPP}
    \end{equation}
A complete Bayesian model requires also the specification of a prior $p(\mu, \sigma, \xi)$, to obtain the posterior $p(\mu, \sigma, \xi \mid \boldsymbol{x}, n_u)$ using Bayes' theorem, $p(\mu, \sigma, \xi \mid \boldsymbol{x}, n_u) \propto p(\mu, \sigma, \xi)L(\boldsymbol{x}, n_u \mid \mu, \sigma, \xi)$. 
This posterior summarizes the information on the parameters after observations, and can be used to extract point estimators, build credible intervals, or write the probability of a new observation $\Tilde{x}$ given data $\boldsymbol{x}$ using the posterior predictive: 
\begin{equation}
    p(\Tilde{x} \mid \boldsymbol{x}, n_u) = \int p(\Tilde{x} \mid \boldsymbol{\theta})p(\boldsymbol{\theta} \mid \boldsymbol{x}, n_u) \dd\boldsymbol{\theta}, \qquad \boldsymbol{\theta} = (\mu, \sigma, \xi).
    \label{eq:post_pred_def}
\end{equation}
These quantities of interest are rarely explicit, and are often derived by sampling approaches. 
A recent survey of extreme value softwares \citep{belzile2022modeler} contains a Bayesian section, and a comparison with frequentist methods.
In the general Bayesian case, an overview of the Bayesian workflow is given in \cite{gelman2020bayesian}, and we focus here on the particular step of reparameterization for the likelihood $L(\boldsymbol{x}, n_u \mid \mu, \sigma, \xi)$ in the case where Markov chain Monte Carlo (MCMC) methods are used to approximate the posterior distribution.

\subsection{Reparameterization}
\label{subsec:reparam}

Although the choice of parameterization of a statistical model does not alter the model \textit{per se}, it does reshape its geometry, which in turn may impact \cor{computational aspects of sampling algorithms} such as efficiency or accuracy. % This is the case for Bayesian inference, and especially in MCMC strategies.
For these methods, a crucial complication for chain convergence is parameter correlation.
This notion of correlation between parameters can be associated with a notion of \cor{asymptotic orthogonality, leading} to independence of posterior components.

\paragraph{Parameterization and \cor{Bayesian inference}}

It has been known for several decades that parameterization is crucial for  good mixing of MCMC chains, especially when the correlation between the coordinates \cor{is large}.
See \citet[Chapter~6]{gilks1995markov} for a great introduction for Gibbs sampling and Metropolis--Hastings algorithm. %For Gibbs, highly dependent components can lead to iterations concentrated to each others, which causes slow mixing as a lot of iterations are therefore required to explore the parameter space. An illustration is given by \cite{hills1992parameterization} with a bivariate Gaussian distribution, where the convergence rate is explicit and is affected by the correlation between the two components.
More general computations are conducted by \cite{Roberts1997} in the normal case, %as the dependence between coordinates can easily be modeled with correlation. However, 
but this convergence rate is less explicit in the general case, see for example \cite{roberts1994geometric}.
For Metropolis--Hastings, if the structure of the kernel is not similar to the one of the target density (which is a typical case if there is a complex dependence between parameters), then too many candidates generated by the kernel are rejected and the same problem as for Gibbs sampling occurs.
For more recent MCMC algorithms such as Hamiltonian Monte Carlo \citep[HMC,][]{neal2011mcmc} and its variant NUTS \citep{hoffman2014no}, \cite{Betancourt2015} gives an example of the benefit of reparameterization for hierarchical models. % Another one can be found in \cite{Vehtari2021Rhat} with Cauchy likelihood, where the issue is not due to correlations but rather to the Cauchy distribution tail-heaviness, that hinders exploration by Hamiltonian dynamics.
More generally, \cite{Betancourt2019} studies reparameterization from a geometric perspective, in order to show its equivalence with adapted versions of HMC on Riemannian manifolds.

Due to the difficulty of obtaining general results on reparameterization and MCMC convergence, a significant part of the research focuses on specific models, such as hierarchical models \citep{Papaspiliopoulos2003, Browne2009}, linear regression \citep{gilks1995markov}, or mixed models  \citep{Gelfand1995, gelfand1996cient}. % An overview of parameterization methods is given in \cite{Gelman2004} in the case of data augmentation and parameters expansion. In addition to improving MCMC convergence,  \cite{Gelman2004} shows that a good parameterization may also help to interpret model parameters. 

For extreme value models, \cite{Diebolt2005} uses a continuous mixture of exponential distributions in the GPD case.
\cite{Opitz2018} also suggests to use the median instead of the usual scale parameter to reduce correlation for Integrated Nested Laplace approximation (INLA).
An alternative Monte Carlo algorithm, the ratio-of-uniforms method, is also implemented for extreme value models in the \texttt{revdbayes} package \citep{revdbayes}. 
The influence of parameterization is also considered in this framework as the acceptance rate can be altered because of correlated parameters (see \ref{subsec:ration_uniform}). %\notcor{Je pense qu'il faut reduire ce paragraphe en citant l'annexe.}
Parameter transformations are also studied in order to make likelihood-based inference suitable in the high-dimensional case in \cite{johannesson2022approximate}.
\corr{Finally, \cite{belzile2022modeler} proposes a reparameterization trick that can be used to obtain a suitable initial value for optimization routines.}

\paragraph{Orthogonal parameterization}

As seen before, reducing dependence between coordinates is desirable for MCMC \cor{methods}.
 \cor{Dependence can be characterized using} asymptotic covariance and the notion of orthogonality according to \cite{Jeffreys61}: parameters are said to be orthogonal when the Fisher information is diagonal. 
\cor{From} this definition, having orthogonal parameters leads to asymptotic posterior independence when a Bernstein--von Mises theorem holds \citep[e.g.,][Chapter~10]{van2000asymptotic}.
\cor{However,} the problem of finding an orthogonal parameterization is seldom feasible when there are more than three parameters, since the number of equations is then greater than the number of unknown variables.
In the case of three parameters, there are as many equations as there are unknowns, but the non linear system does not necessarily lead to a solution \citep{Huzurbazar1950}.

The main use of orthogonal parameterization is to make parameters of interest independent of nuisance parameters \citep{Cox1987}.
Other definitions of orthogonality are also proposed to be more adapted to the inferential context \citep{Tibshirani1994} or to ensure consistency of the parameter of interest \citep{Woutersen2011}.
For Bayesian inference, \cite{Tibshirani1994} compares different definitions and suggests a strong assumption of normality for the posterior. 
In the following, we keep the most popular definition of orthogonality due to \cite{Jeffreys61}, as we are not interested in properties associated with the estimation of a given parameter of interest, but rather on the dependence structure between parameters.
However, up to our knowledge, there is no clear evidence in the literature of a direct link between parameter orthogonality and mixing properties of the corresponding MCMC chains, such as a better convergence rate.
In Section~\ref{sec:exp}, we bring some empirical evidence on the interest of orthogonality in extreme value models.

\subsection{Contributions and outline}

In this paper, we study the benefits of reparameterization for the Poisson process characterization of extremes in a Bayesian context.
In particular, \corr{it is shown} that the orthogonal parameterization is useful for several reasons: we argue in Section~\ref{subsec:sharkey} that it improves the performance of MCMC algorithms in terms of convergence, and we show in Section~\ref{sec:priors} that it also facilitates the derivation of priors such as Jeffreys and an informative variant on the shape parameter using penalized complexity (PC) priors \citep{Simpson2017}. 
These results are then illustrated by experiments in Section~\ref{sec:exp}, first on simulations to compare the different parameterizations, and second on a real dataset of the Garonne river flow.
Proofs as well as additional experiments are provided in the Appendix, and the code corresponding to the experiments is available online.\footnotemark{}
\footnotetext{\url{https://github.com/TheoMoins/ExtremesPyMC}}

\section{Reaching orthogonality for extreme Poisson process}
\label{subsec:sharkey}

An attempt to reparametrize the Poisson process for extremes in order to improve MCMC convergence already exists in the literature \citep{Sharkey2017}, but has several limitations that \corr{are detailed} here. 
Instead, we suggest to use the fully orthogonal parameterization of \cite{Chavez-Demoulin2005}.

\paragraph{Near-orthogonality with hyperparameter tuning}

Based on the relationship between parameters given in Equation~(\ref{eq:mu_m}), \cite{Sharkey2017} suggests to change the scaling factor $m$ before using Metropolis--Hastings algorithm
%to \sg{minimize} the \sg{absolute} asymptotic covariance between parameters,
in order to optimize MCMC convergence. 
To this aim, they minimize the non-diagonal elements of the inverse Fisher information matrix corresponding to asymptotic covariances and then retrieved the parameters corresponding to the initial number of blocks from Equation~(\ref{eq:mu_m}).
As the calculations cannot be achieved explicitly, the authors found empirically that the values $m_1$ and $m_2$ that cancel respectively the asymptotic covariances $\text{ACov}(\mu, \sigma)$ and $\text{ACov}(\sigma, \xi)$ are such that any $m \in [m_1, m_2]$ improves the MCMC convergence.
Approximations of $m_1$ and $m_2$ are then given as functions of $\xi$, and therefore a preliminary estimation of $\xi$ (typically obtained using maximum likelihood estimation) is required to obtain $\hat{m}_1(\xi)$ and $\hat{m}_2(\xi)$, and to choose a value in this interval before running an MCMC with the right choice of $m$.
Despite leading to significant improvement of the convergence of Markov Chains, this method suffers from several limitations.
First, preliminary estimation of the shape parameter $\xi$ is required, \cor{to compute} $\hat{m}_1(\xi)$ and $\hat{m}_2(\xi)$ and \cor{choose} a value in the corresponding interval, which adds complexity and computational burden on the overall framework.
Moreover, it also affects the accuracy of orthogonalization, as the expressions of $m_1$ and $m_2$ are found empirically, then are approximated by $\hat{m}_1(\xi)$ and $\hat{m}_2(\xi)$, and finally computed at $\hat{\xi}$ which adds a new source of uncertainty.
One way to lighten the method would be to suggest a simpler choice of $m$, for example $m=n_u$,  which leads to a satisfactory behaviour as noticed by \cite{Wadsworth2010}.
However, \corr{it is shown} in \ref{sec:sharkey} that this choice presents some flaws and does not bring any general guarantee of orthogonality.

\paragraph{Orthogonal parameterization}

\cor{Alternatively}, there exists a parameterization of the Poisson process that leads to orthogonality. 
Suggested by \cite{Chavez-Demoulin2005}, it consists of the change of variable
\begin{equation}
    \label{eq:reparam}
    (r, \nu, \xi) = \left(m\left(1+\xi\left(\frac{u-\mu}{\sigma}\right)\right)^{-1/\xi}, (1+ \xi)(\sigma + \xi(u-\mu)), \xi\right),
\end{equation}
\cor{while the inverse transformation is
\begin{equation*}
    (\mu, \sigma, \xi) = \left(u - \frac{\nu}{\xi(1+\xi)}\left(1-\left(\frac{r}{m}\right)^{\xi}\right), \frac{\nu}{(1+ \xi)}\left(\frac{r}{m}\right)^\xi, \xi\right).
\end{equation*}}
With this parameterization, the likelihood \cor{is}
\begin{equation}
        L(\boldsymbol{x}, n_u \mid r, \nu, \xi) = e^{-r}\left(\frac{r}{m}\right)^{n_u} \left(\frac{\nu}{1+\xi}\right)^{-n_u}
        \prod_{i=1}^{n_u} \left(1+ \frac{\xi(1+\xi)}{\nu}(x_i-u) \right)^{-1-1/\xi}.
        \label{eq:orthogonal_fisher}
\end{equation}
Under this form, we can directly see that $r$ is orthogonal to $\nu$ and $\xi$, as the likelihood factorizes with respect to $r$ and $(\nu,\xi)$.
Parameter $r \geq 0$ represents the intensity of the Poisson process, which is the expected number of exceedances, while the two other ones can be seen as an orthogonal parameterization of the GPD with scale $\Tilde{\sigma}_u = \sigma + \xi(u-\mu)$ and shape $\xi$.
%The motivations of \cite{Chavez-Demoulin2005} for this transformation are different from ours, as it is used for a generalised linear model for extremes. 
Under this parameterization and provided $\xi > -{1}/{2}$, the Fisher information matrix $\mathcal{I}(r, \nu, \xi)$ is
\begin{equation}
    \mathcal{I}(r, \nu, \xi) = \text{diag}\left(\frac{1}{r},  \frac{r}{\nu^2(1+2\xi)}, \frac{r}{(1+\xi)^2}\right),
    \label{eq:fisher_reparam_matrix}
\end{equation}
where $\text{diag}(\boldsymbol{u})$ denotes the diagonal matrix with diagonal equal to vector $\boldsymbol{u}$. 
Calculations are provided in \ref{sec:proof_prior}.
Therefore, the orthogonal parameterization of \cite{Chavez-Demoulin2005} is more adapted than the tuning of $m$ since it directly yields the optimal solution sought by \cite{Sharkey2017}.
Moreover, it is obtained without recourse to any optimization procedure or approximation. 
Finally, by plugging $(r, \nu)$ into Equation~(\ref{eq:mu_m}), \corr{one} can show that the invariance property with respect to $m$ holds for the three parameters, and so the parameterization is independent of the choice of $m$.

\paragraph{Generalisation to covariates}
\corr{
The inclusion of covariates in a model holds both theoretical and practical significance, as it enables the incorporation of factors such as temporal trends. 
A notable advantage of this approach is that if the parameters are orthogonal and each of them depends on distinct parameters, then these parameters will also be orthogonal to one another.
To elaborate further, let $\boldsymbol{C}_i$ represent the set of covariates associated with the observation $x_i$. 
In the most comprehensive scenario, where all relevant covariates are considered, the model can be expressed as follows:
$$
    \begin{cases}
    r_i = f_r(\boldsymbol{\theta}_r, \boldsymbol{C}_i),\\
    \nu_i = f_\nu(\boldsymbol{\theta}_\nu, \boldsymbol{C}_i),\\
    \xi_i = f_\xi(\boldsymbol{\theta}_\xi, \boldsymbol{C}_i),
    \end{cases}
$$
where $\boldsymbol{\theta}_r$, $\boldsymbol{\theta}_\nu$ and $\boldsymbol{\theta}_\xi$ are three vectors of parameters.
The log-likelihood can be written as
$$
\sum_{i=1}^n \ell_i(r_i,\nu_i,\xi_i)=\sum_{i=1}^n \ell_i(f_r(\boldsymbol{\theta}_r, \boldsymbol{C}_i), f_\nu(\boldsymbol{\theta}_\nu, \boldsymbol{C}_i), f_\xi(\boldsymbol{\theta}_\xi, \boldsymbol{C}_i)).
$$
By leveraging the property that no parameters are shared, along with the chain rule, one can derive the following expression for the Fisher information associated with the $j$th coordinate $\theta_{\nu,j}$ of $\boldsymbol{\theta}_\nu$ and the $k$th coordinate $\theta_{\xi,k}$ of $\boldsymbol{\theta}_\xi$:
\begin{equation*}
    \mathbb{E}\left(-\frac{\partial^2 \ell_i}{\partial \theta_{\nu,j} \partial \theta_{\xi,k}}\right)
    = \frac{\partial f_\nu}{\partial \theta_{\nu,j}}\frac{\partial f_\xi}{\partial \theta_{\xi,k}}
    \mathbb{E}\left(-\frac{\partial^2  \ell_i }{\partial \nu_i \partial \xi_i}\right) 
    = 0.
\end{equation*}
As a result, every parameter in $\boldsymbol{\theta}_\nu$ is orthogonal to all parameters in $\boldsymbol{\theta}_\xi$. 
Similar calculations reveal that $\boldsymbol{\theta}_r$ is orthogonal to both $\boldsymbol{\theta}_\nu$ and $\boldsymbol{\theta}_\xi$. 
Hence, when covariates are defined based on orthogonal parameters, it leads to block-wise orthogonality, with the associated Fisher information matrix comprising three blocks in this case.
}

\section{Priors invariant to reparameterization}
\label{sec:priors}

In the case where no external information is available about the parameters, the choice of the prior distribution should be made with caution.
Typically, the term ``uninformative prior'' or ``objective prior'' can be misleading, as it refers to priors used when one does not have preliminary information, but the prior itself does contain information.
\cor{As an example, a flat prior over the range of possible values is only flat for a given parameterization.
After a change of variable, a uniform prior does not necessarily remain uniform \citep[\corr{e.g., }][Chapter~3]{robert2007bayesian}.}
This problem is all the more serious as our study which deals with reparametrization:
% for the Poisson process model, how to justify a uniform prior over $(\mu, \log \sigma, \xi)$, $(r, \log \frac{\nu}{1+\xi}, \xi)$, or something else? Even in the informative case, two experts that provide equivalent quantities on two parameterizations should expect the same result in the end.
here, we derive two priors that enjoy the property of being invariant with respect to reparameterization.

\subsection{Jeffreys prior}
\label{subsec:jeffreys}

Jeffreys prior \citep{jeffreys1946} is built with the aim of invariance: if $\mathcal{I}(\boldsymbol{\theta})$ denotes the Fisher information matrix associated with parameters $\boldsymbol{\theta}$, it is defined as 
\begin{equation}\label{eq:jeffreys}
    p_\mathrm{J}(\boldsymbol{\theta}) \propto \sqrt{\det \mathcal{I}(\boldsymbol{\theta})}.
\end{equation}
Under this prior, one can show that a reparameterization $\boldsymbol{\phi} = h(\boldsymbol{\theta})$ \cor{with $h$ a continuously differentiable function} yields $p_\mathrm{J}(\boldsymbol{\phi}) \propto \sqrt{\det \mathcal{I}(\boldsymbol{\phi})}$.
This prior is computed for the GPD by \citet{EugeniaCastellanos2007} and for the GEV under a modified version where $p_\mathrm{J}(\mu, \sigma, \xi) \propto \sqrt{\det \mathcal{I}(\sigma, \xi)}$ by \cite{kotz2000extreme}.
Up to our knowledge, Jeffreys prior has never been computed for the Poisson process characterization of extremes.
The orthogonalization done in Equation~(\ref{eq:reparam}) directly provides Jeffreys prior with respect to $(r, \nu, \xi)$:
\begin{Prop}
\label{first-prop}
    Jeffreys prior associated with a Poisson process for extremes with parameters $(r, \nu, \xi)$ from Equation~(\ref{eq:reparam}) exists provided $\xi>-1/2$, and is given by
    \begin{equation}
        p_\mathrm{J}(r, \nu, \xi) \propto \frac{r^{1/2}}{\nu (1+\xi) (1 + 2 \xi)^{1/2}}.
        \label{eq:jeffreysPPortho}
    \end{equation}
\end{Prop}
Moreover, the invariance to reparameterization property \cor{directly} provides  the expression of Jeffreys prior on $(\mu, \sigma, \xi)$.
\begin{Cor}
Jeffreys prior associated with a Poisson process for extremes with original parameters $(\mu, \sigma, \xi)$ exists provided $\xi>-1/2$, and can be written as
    \begin{equation}
            p_{J}(\mu, \sigma, \xi) \propto
            \frac{\left(1 + \xi \left(\frac{u-\mu}{\sigma}\right)\right)^{-\frac{3}{2\xi}-1}}
            {\sigma^2 (1+\xi) (1 + 2 \xi)^{{1}/{2}}}.
            \label{eq:jeffreysPP}
    \end{equation}
\end{Cor}
This prior cannot be defined for $\xi \leq -1/2$, as it corresponds to a case where the Fisher information matrix is infinite. 
However, this assumption is not too restrictive as the great majority of models of interest belong to a maximum domain of attraction with $\xi \in (-1/2, 1/2)$.
Note that this prior, similarly to the uniform one, is improper in the sense that the integral over the range of parameters is infinite.
Consequently, it is necessary to check whether the posterior is proper or not to be able to use it.
\cite{EugeniaCastellanos2007} shows that the posterior is proper when using Jeffreys prior in the GPD case, while \cite{Northrop2016} shows that it is never the case with GEV likelihood.
For the Poisson process, we show the following result:
\begin{Prop}
    \label{prop:jeffreys_proper}
    Jeffreys prior for a Poisson process for extremes yields a proper posterior distribution, as long as $\xi>-1/2$.
\end{Prop}
A proof is provided in \ref{sec:proof_prior}.

\subsection{Penalized complexity prior for the shape parameter}
\label{subsec:pc_prior}

\begin{figure}[ht]
    \begin{tabular}{cc}
        \hspace{-0cm}\raisebox{-.55\height}{\includegraphics[trim={0cm 0cm 0cm 0cm},clip,width =0.44\textwidth]{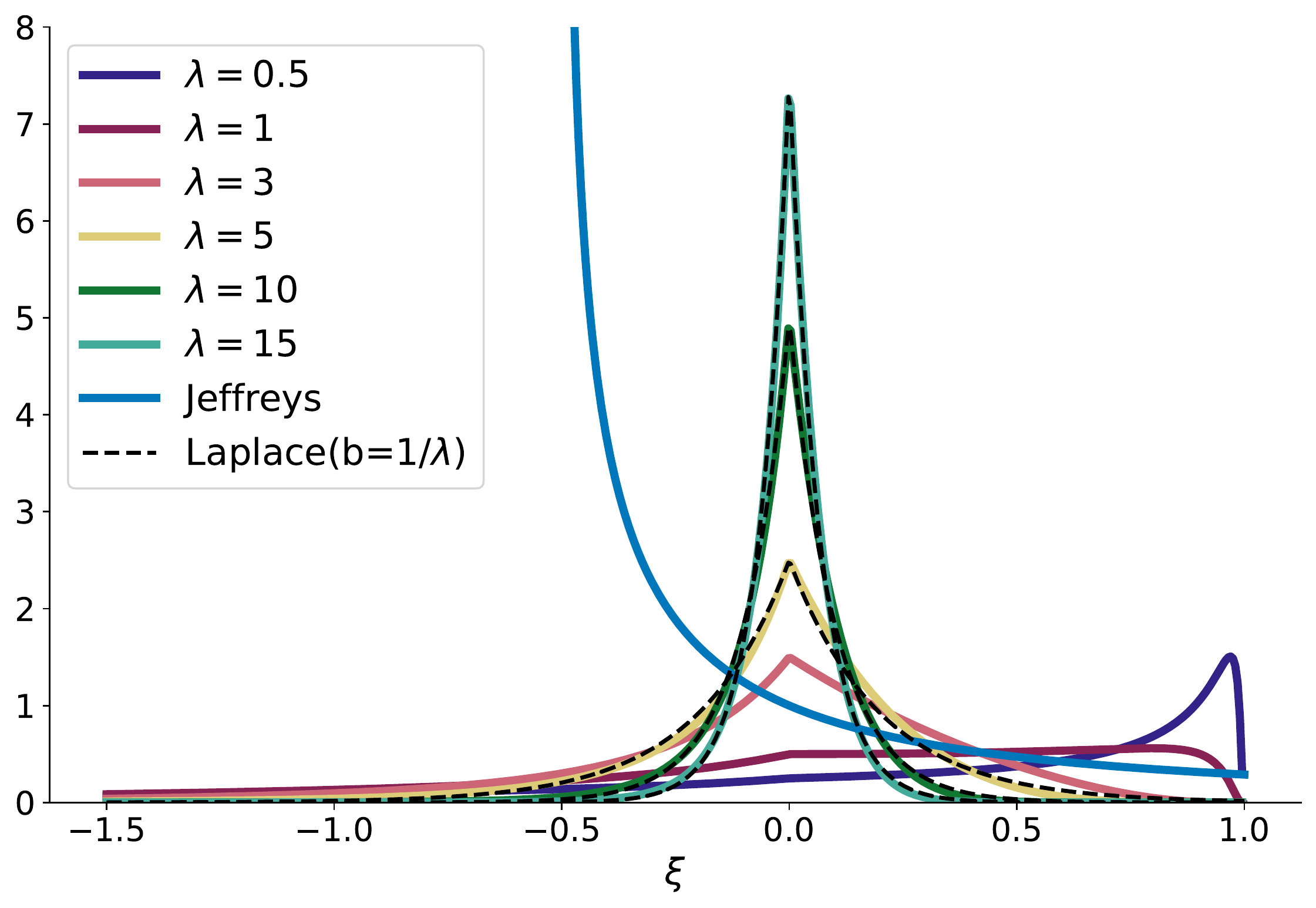}} & 
        \begin{tabular}{ccc}
            \toprule
            $\lambda$ & $\text{I}_\mathrm{PC}(0.95 \mid \lambda)$ & $\text{I}_\text{Lap}(0.95 \mid \lambda)$ \\
            \midrule
            $0.5$ & $[-36.8, 0.97]$ & $[-6.00, 6.00]$ \\
            $1$ & $[-9.88, 0.90]$ & $[-2.99, 2.99]$ \\
            $3$ & $[-1.61, 0.61]$ & $[-0.99, 0.99]$ \\
            $5$ & $[-0.80, 0.44]$ & $[-0.59, 0.59]$ \\
            $10$ & $[-0.34, 0.25]$ & $[-0.29, 0.29]$ \\
            $15$ & $[-0.22, 0.18]$ & $[-0.19, 0.19]$ \\
            \bottomrule
        \end{tabular}
    \end{tabular}
    \caption{Left panel: examples of PC priors $p_\mathrm{PC}(\cdot \mid \lambda)$ with $\lambda$ \cor{ranging} from $0.5$ to $15$, and Jeffreys prior (blue curve) represented for fixed values of $(\mu, \sigma) = (0,1)$. The black dashed lines represent Laplace distributions with scale parameter equal to $1/\lambda$, for $\lambda\in\{5,10,15\}$. Note that Laplace distributions $p_{\mathcal{L}}(\cdot  \mid 1/\lambda)$ approximate well $p_{\mathrm{PC}}(\cdot \mid \lambda)$ when $\lambda \geq 10$. 
    Right panel: credible intervals at $95\%$ for PC and Laplace priors, resp. $\text{I}_\mathrm{PC}(0.95 \mid \lambda)$ and $\text{I}_\text{Lap}(0.95 \mid \lambda)$. 
    }
    \label{fig:pc_prior}
\end{figure}

The shape parameter $\xi$ plays a crucial role in the \cor{inference}, as it tunes the heaviness of the tail distribution: it is heavy if $\xi > 0$, light if $\xi = 0$ and finite (\textit{i.e.} with a finite right end-point) if $\xi < 0$.
The case $\xi = 0$ can be seen as a simpler model with an exponential decrease of the survival function, where the GPD cdf in Equation~(\ref{eq:gpd_distribution}) simplifies to an exponential distribution.
This concentration of an entire maximum domain of attraction at a single value of $\xi$ complicates the study, \cor{since} it is for example difficult to distinguish heavy tails with low $\xi$ \cor{from} light tails \citep{Stephenson2004}.
However, this change of regime can have significant consequences when it comes to extrapolation.
It should also be noted that a vast majority of datasets have distribution with \cor{$|\xi|\leq1/2$}.
It is therefore \cor{natural, even in a non-informative framework, to penalize high values of $|\xi|$}.
One way to do this is to use penalized complexity (PC) priors \citep{Simpson2017}: the idea is to consider a prior that penalizes exponentially the distance between a model $p_\xi := p(\cdot \mid \xi)$ with a given $\xi$ and the baseline $p_0$ with $\xi = 0$.
The general formula is
\begin{equation*}
    p_{\text{PC}}(\xi \mid \lambda) = \lambda \exp(-\lambda d(\xi)) \left|\frac{\partial d(\xi)}{\partial\xi}\right|, 
\end{equation*}
with $\lambda > 0$, $d(\xi) = \sqrt{2\text{KL}\left(p_\xi || p_0\right)}$ and KL$(p_\xi || p_0)$ the Kullback--Leibler divergence between $p_\xi$ and $p_0$: $\text{KL}(p_\xi || p_0)= \int p_\xi(x) \log \left(p_\xi(x) / p_0(x) \right) \, \dd x$.
Parameter $\lambda$ acts as a scaling parameter and controls the range of acceptable values for $\xi$.
This prior has the advantage of being proper and invariant to reparameterization on $\xi$.
The computation with GPD has already been done by \cite{Opitz2018} for the case $\xi \geq 0$: the authors prove that $d(\xi)$ is finite only if $\xi < 1$, and \cor{is} $d(\xi) = \sqrt{2}\xi/\sqrt{1-\xi}$ for $0 \leq \xi < 1$.
Then, they show that it can be approximated by an exponential distribution on $\xi$ in the case $\xi \to 0$, when $\lambda$ can be taken sufficiently large and to favor $\xi = 0$.
A first observation is that routine calculations extend this definition both to negative values of $\xi$, and to the Poisson process characterization where the density of observation is also GPD.
\begin{Prop}
   The PC prior associated with a Poisson process for extremes exists for any $\xi < 1$ and \cor{is}
    \begin{equation}
        p_\mathrm{PC}(\xi \mid \lambda) = \frac{\lambda}{2}\left(\frac{1 - \xi/2}{(1-\xi)^{3/2}}\right) \exp\left(-\lambda \frac{|\xi|}{\sqrt{1-\xi}}\right).
        \label{eq:pc_prior}
    \end{equation}
\end{Prop}
This prior is plotted for several values of $\lambda$ in Figure~\ref{fig:pc_prior}.
\cor{As observed by} \cite{Opitz2018}, \cor{the PC prior} is very similar to a Laplace$(0,1/\lambda)$ when $\lambda$ is \cor{large enough so that} the peaks at 0 dominates over the endpoint at 1.
\cor{In the case where $\lambda$ is small (typically $\lambda \leq 1$), an asymptote appears at the upper bound $\xi = 1$ which could have an undesirable influence in the posterior distribution. 
Thus, for the least informative case, a value of $\lambda = 1$ is sufficiently small as it does not favor values close to $0$ nor those close to 1.}
In the case when $0$ is favoured with a high $\lambda$ and the true value of $\xi$ differs from $0$, the estimation may be altered compared to the uninformative case: see \ref{sec:replications_mse} for an analysis on simulated data.
For the two other parameters, one can consider Jeffreys' rule on $(r, \nu)$ in order to obtain a non-informative prior for $(r, \nu)$ while keeping invariance to reparameterization property. 
$\xi$ is therefore considered \textit{a priori} independent of $(r, \nu)$. 
\cor{In view of} the Fisher information matrix in Equation~(\ref{eq:fisher_reparam_matrix}), we obtain $p_\mathrm{J}(r, \nu) \propto 1/\nu$.
Similarly to Jeffreys prior in Section~\ref{subsec:jeffreys}, the resulting prior is improper but the following proposition \corr{can be shown}:
\begin{Prop}
    \label{prop:pc_proper}
    The prior defined as $p(r, \nu, \xi) \propto p_\mathrm{PC}(\xi)p_\mathrm{J}(r, \nu) \propto p_\mathrm{PC}(\xi)/\nu$ for the Poisson process for extremes yields a proper posterior distribution.
\end{Prop}
The proof, detailed in \ref{sec:proof_prior}, relies on a result of \cite{Northrop2016}. 
Note that this result still holds if $p_\mathrm{PC}(\xi)$ is replaced by its approximation through the Laplace distribution.

\section{Experiments}
\label{sec:exp}

The benefits of the orthogonal reparameterization in the Poisson process model \corr{are illustrated here} on simulations and a real environmental dataset. \ref{sec:additional_exp} contains additional experiments, notably using Hamiltonian Monte Carlo (HMC) instead of MCMC (\ref{subsec:NUTS}), under various maximum domains of attraction (\ref{subsec:simul_appendix}), in other models than the Poisson process model, \cor{namely} the GPD and GEV ones (\ref{subsec:gev_gpd}), {using the ratio-of-uniforms algorithm instead of MCMC (\ref{subsec:ration_uniform})} and finally with replications and comparison with maximum likelihood (\ref{sec:replications_mse}).
All experiments are done using PyMC3 library \citep{PyMC3}, and the corresponding code is available online\footnotemark[\value{footnote}]{}.

\subsection{Simulations with the Poisson process model}
\label{subsec:simulations}

\paragraph{Data generation}
We start by comparing the different parameterizations on exceedances generated with the Poisson process model described in Section~\ref{subsec:bayes_ext_intro}.
For a given value of $(\mu, \sigma, \xi)$ and hyperparameters $(u, m)$, the data generation proceeds in two steps: 
first, a number of events $n_u$ is simulated using a Poisson distribution with parameter $\Lambda(I_u)$ as defined in Section~\ref{subsec:bayes_ext_intro}.
Then, for each point $i \in \{1, \ldots, n_u\}$, the position $x_i$ knowing that $x_i \in I_u$ is sampled from a GPD with parameters $(u, \Tilde{\sigma}, \xi)$, with $\Tilde{\sigma} = \sigma + \xi (u-\mu)$.
An example with $(m, u, \mu, \sigma, \xi) = (40, 30, 50, 15, -0.25)$ is detailed here, \cor{leading} to an expected number of observations $\Lambda(I_u) \approx 126$. 

\paragraph{Experimental setup}
For MCMC hyper-parameters such as number of chains, burn-in period per chain or initialization, we keep the default values suggested in the PyMC3 library: 
\cor{four chains (which corresponds to our number of cores) with $1\,000$ iterations each, and a burn-in period of $1\,000$ iterations.}
In addition to these choices, this library offers the possibility to choose among different sampling methods, such as the traditional Metropolis--Hastings algorithm, but also more modern MCMC algorithms like Hamiltonian Monte Carlo \citep[HMC,][]{neal2011mcmc}, or the No-U-Turn sampler \citep[NUTS,][]{hoffman2014no} which is the default choice in PyMC3.
We choose to compare the different reparameterizations on Metropolis--Hastings draws (after burn-in), and the behaviour on NUTS is also investigated \ref{subsec:NUTS}.
We show that $1\,000$ iterations are sufficient for the chains to converge when the parameterization is well chosen. 
However, note that the algorithm only takes a few seconds to run, so this number of iterations can easily be increased for real data  applications, as done in Section~\ref{subsec:real_data}.
Finally, \corr{Jeffreys prior (computed in Section~\ref{subsec:jeffreys}) is choosen} for all configurations, but experiments have shown similar results with the PC prior of Section~\ref{subsec:pc_prior}.

\paragraph{Convergence diagnostic}
Our aim is to discriminate the different parameterizations according to the rate of convergence of the MCMC chains to their target. Different indicators exist to \cor{assess} the quality of MCMC approximation. 
First, given a finite number of samples, autocorrelation plots as \cor{functions} of lag measure how good the posterior approximation is, as the dependence between the \cor{chain} elements reduces the effective information available for inference. 
To measure this, a common practice relies on the effective sample size, defined as $\text{ESS} = MN(1 + 2 \sum_{t=1}^{\infty} \rho_t)^{-1}$, 
with $M$ the number of chains of size $N$, and $\rho_t$ the autocorrelation at lag $t$.
\cor{The ESS} corresponds to an equivalent number of independent draws, and so quantifies the amount of effective data for estimation \citep[\corr{e.g., }][Section~11.5]{gelman2013bayesian}.
Here, the evolution of ESS with the number of draws is reported \cor{for each configuration}.
To complete the diagnostic, the potential scale reduction factor (commonly denoted by $\hat{R}$) also aims at bringing an indication about the state of convergence by computing the ratio of two estimators of the posterior variance.
Generally $\hat{R} \geq 1$, and if it is greater than a given threshold, a convergence issue is raised.
We use here a refinement of $\hat{R}$ named $\hat{R}_\infty$ \citep{moinsRhatArxiv}, based on a local version $\hat{R}(x)$ which aims at ensuring the convergence at a given quantile $x$ of the distribution.
Then, $\hat{R}_\infty$ is defined as the supremum of the $\hat{R}(x)$ values: $\hat{R}_\infty := \sup_{x \in\mathbb{R}} \hat{R}(x)$. 
This scalar summary \cor{amounts} to considering the value of $\hat{R}(x)$ associated with the worse quantile approximation by the MCMC chains.

\paragraph{Results}

\begin{figure}
    \includegraphics[trim={0cm 0cm 0cm 0cm},clip,width =1\textwidth]{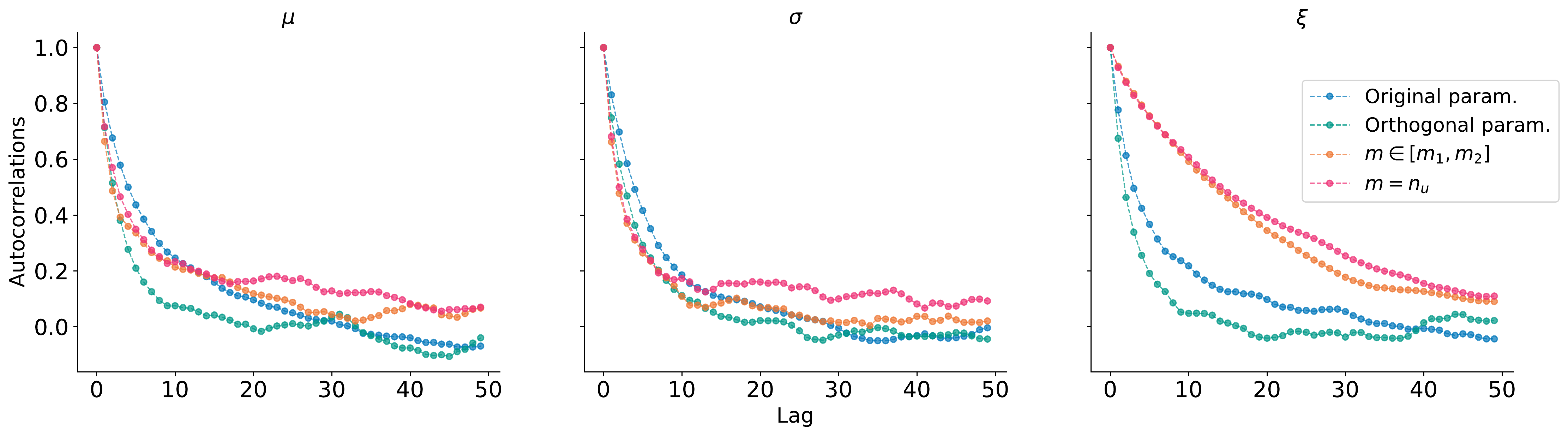}

    \includegraphics[trim={0cm 0cm 0cm 0cm},clip,width =1\textwidth]{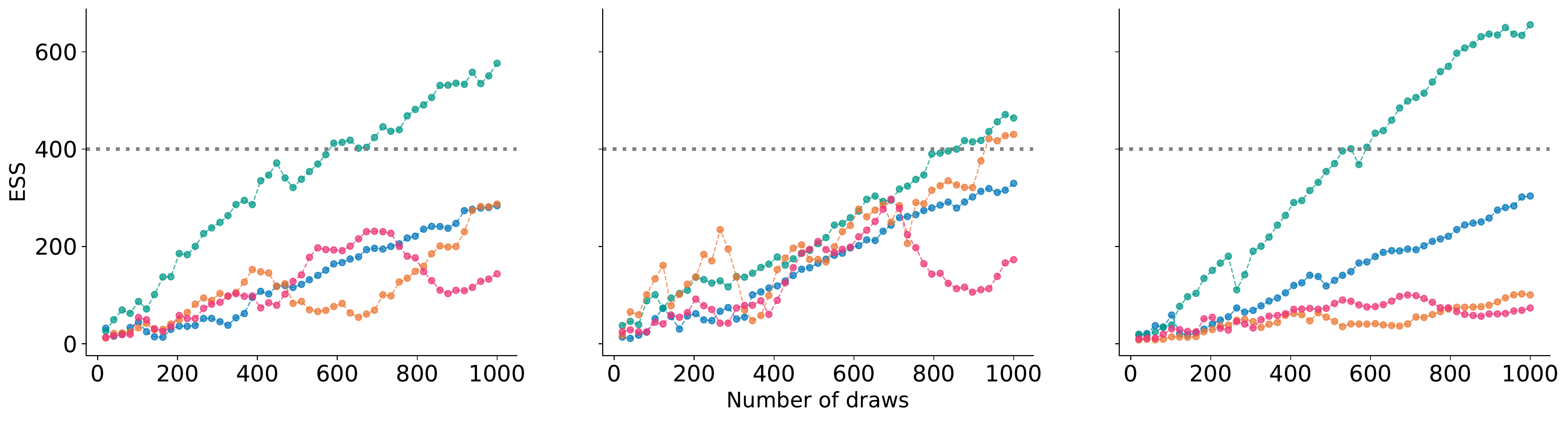}

    \includegraphics[trim={0cm 0cm 0cm 0cm},clip,width =1\textwidth]{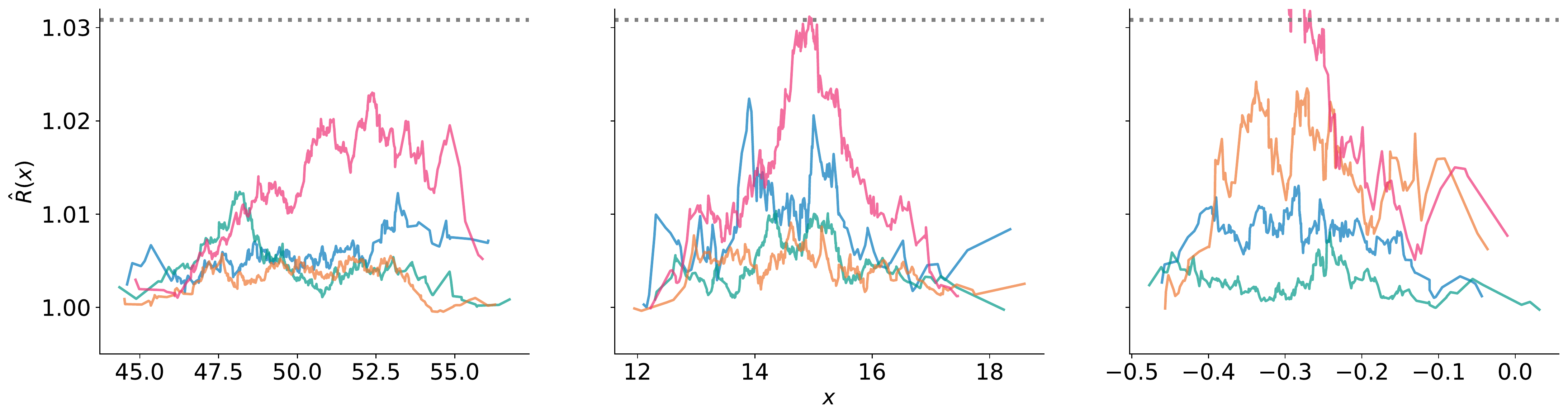}
    \caption{Convergence diagnostic plots for Poisson parameters $(\mu, \sigma, \xi)$ with $\xi < 0$, after $1\,000$ Metropolis--Hastings draws and a burn-in of $1\,000$, for four different parameterizations: the original one (in \textcolor{colorblind1}{blue}), the \cite{Sharkey2017} update with $m \in [\hat{m}_1, \hat{m}_2]$ (in \textcolor{colorblind3}{orange}), the \cite{Wadsworth2010} update with $m = n_u$ (in \textcolor{colorblind4}{magenta}), and the orthogonal parameterization (in \textcolor{colorblind2}{green}).
    Top row: autocorrelations as functions of the lag. 
    Second row: evolution of ESS with the number of draws (the gray line corresponds to value of $400$ recommended in \cite{gelman2013bayesian}).
    Bottom row: $\hat{R}(x)$ as a function of the quantile $x$, with the adapted threshold of $1.03$ \citep{moinsRhatArxiv}.
    Some curves are truncated for visibility purposes, as they are taking much larger values than the threshold.}
    \label{fig:PP_xi_negative}
\end{figure}

Results are reported in Figure~\ref{fig:PP_xi_negative}, with four parameterizations that are compared for MCMC efficiency.
(i) The orthogonal parameterization $(r, \nu, \xi)$ of Equation~(\ref{eq:reparam}), and three triplets $(\mu, \sigma, \xi)$ associated with the following choices of $m$: (ii)  the original $m$ (same as the one used for generation), (iii) $m = n_u$ which is the choice of \cite{Wadsworth2010} and the package \texttt{revdbayes} \citep{revdbayes}, and (iv)  $m \in [m_1, m_2]$ as suggested by \cite{Sharkey2017} (see Section~\ref{subsec:sharkey}).

In order to compare the same quantities, all convergence diagnostics are computed with the original parameterization $(\mu, \sigma, \xi)$ and $m$, consequently after a transformation of the chains for the other parameterizations.
Figure~\ref{fig:PP_xi_negative} \corr{confirms} that the orthogonal parameterization behaves best in the case $\xi < 0$: 
the \cor{Markov chains} have the lowest autocorrelations, the lowest value of $\hat{R}(x)$ for almost all $x$, and this parameterization is the only one that satisfies the recommendation of $\text{ESS}$ lower than $400$ for estimation \citep{gelman2013bayesian}.
Conversely, the two parameterizations that suggest a change for $m$ seem to suffer from a lack of convergence, even more than the original parameterization.
For the cases $\xi > 0$ and $\xi = 0$ detailed in \ref{subsec:simul_appendix}, the orthogonal parameterization is still best, but the behaviour of the three other parameterizations is reversed: the one with no change for $m$ is the one with the largest convergence issues.
Some intuitions about the behaviour of parameterizations that rely on changing $m$, in particular in the case $\xi < 0$, can be found in \ref{sec:sharkey}.
We also refer to \ref{subsec:gev_gpd} for a study of the GPD and GEV cases.
As a conclusion, the orthogonal parameterization is effective in the three maximum domains of attraction, for both Poisson \cor{process} and GPD models.

\subsection{Case study on river flow data}
\label{subsec:real_data}

We apply our framework \cor{to $36\,160$} daily measurements of the Garonne river flow (France), from 1915 to 2013.

\begin{figure}[h]
    \centering
    \includegraphics[trim={0cm 0cm 0cm 0cm},clip,width =0.7\textwidth]{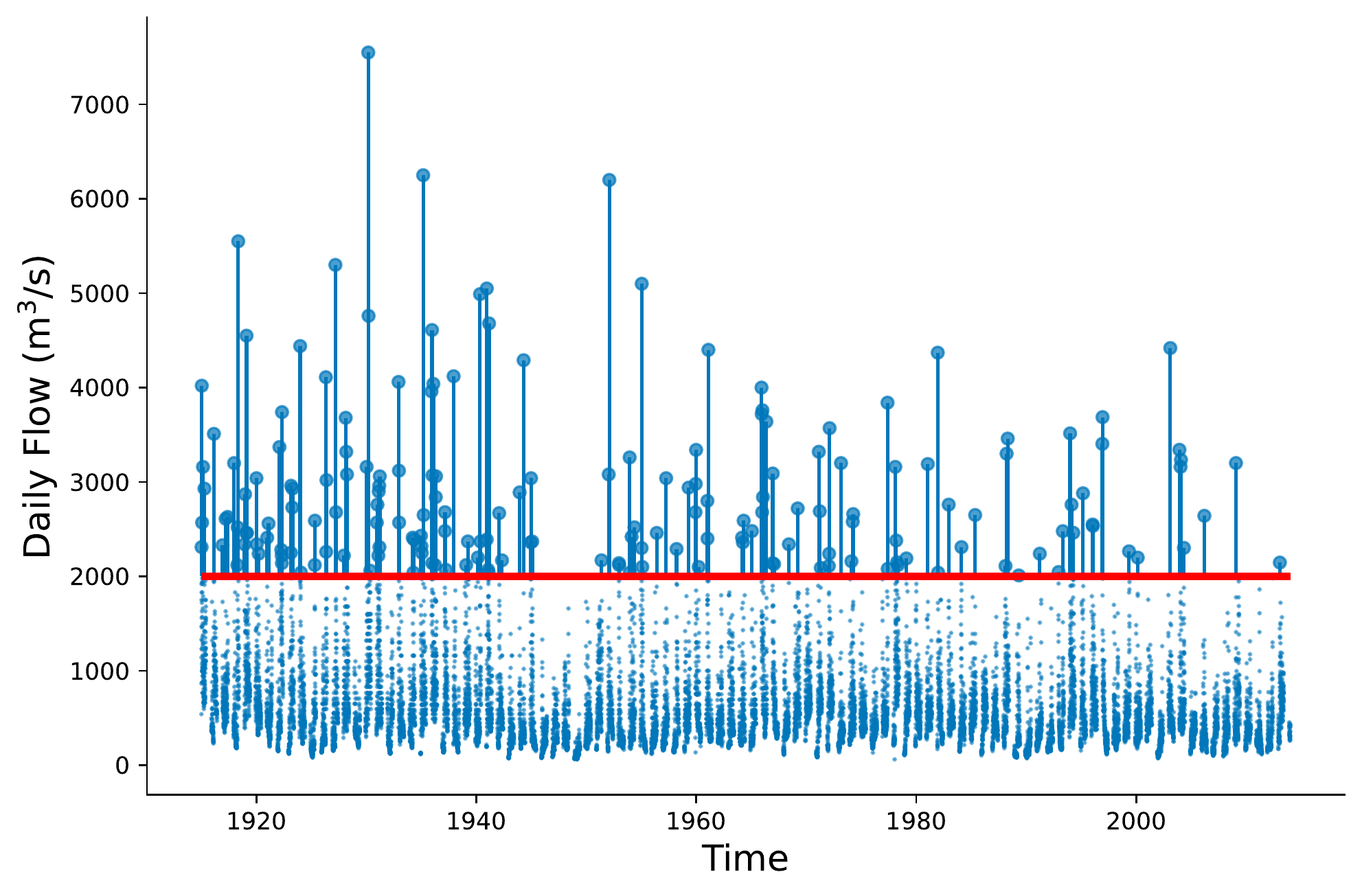}
    \caption{Plot of $n_u=182$ exceedances of the Garonne river flow between 1915 and 2013 above the threshold $u=2\,000$ (represented in red).}
    \label{fig:real_data}
\end{figure}

\paragraph{Preprocessing}
Before selecting a threshold and running a MCMC algorithm, some common preprocessing steps on daily environmental data are required: first because of seasonality, only the rainy season from December to May \corr{is considered}, which reduces the number of observations to $18\,043$. 
The observations are also not independent; an autocorrelation plot suggests a three-day correlation in measurements. 
Therefore, clusters of exceedances of parameter $r=3$ days are considered here, which means that two exceedances that occurred in less than three days are merged as one observation (the largest one in the cluster).
Previous EDF studies \citep[see for instance][]{albert:tel-01971408} agree with traditional threshold elicitation methods \citep[\corr{e.g., }][Chapter~4]{Coles2001} to consider a threshold of $u=2\,000$ m$^3$/s for estimation. 
In the end, $n_u=182$ clusters of exceedances are \corr{obtained and} represented in Figure~\ref{fig:real_data}.
\corr{A temporal trend could be suspected there. A possible way to model such a phenomenon would be to include covariates in the orthogonal parameters, see the last paragraph of Section~\ref{subsec:sharkey}.}

\begin{figure}
    \centering
    \includegraphics[trim={0cm 0cm 0cm 0cm},clip,width =\textwidth]{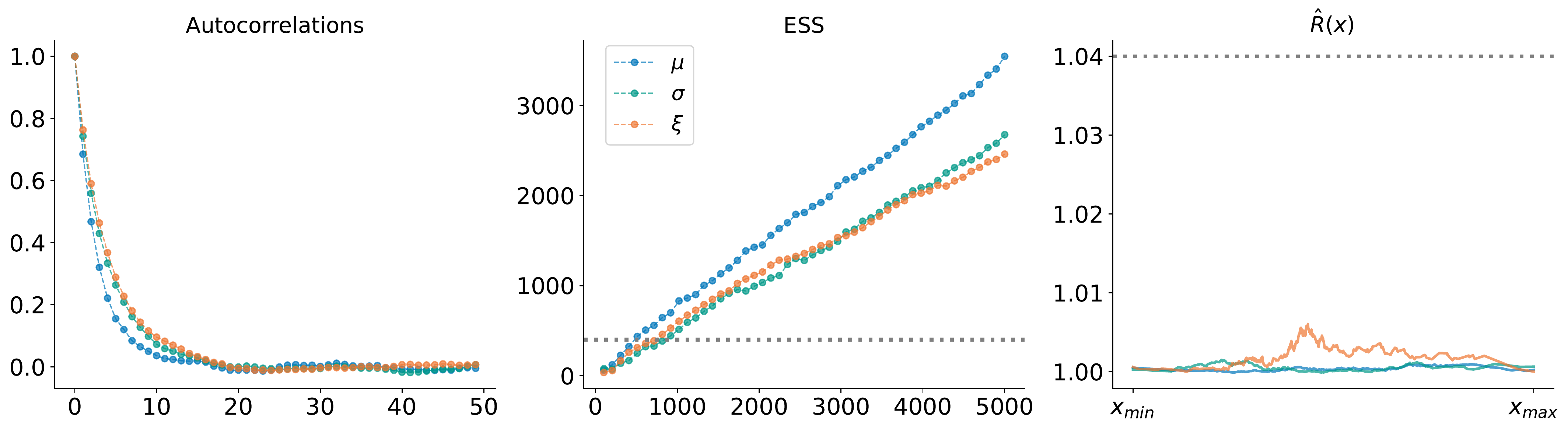}
 
    \caption{Convergence diagnostic plots for Garonne river flow data, after $5\,000$ Metropolis--Hastings draws and a burn-in of $1\,000$.
    Left: autocorrelations as functions of the lag. 
    Middle: evolution of ESS with the number of draws (the gray line corresponds to value of $400$ recommended in \cite{gelman2013bayesian}).
    Right: $\hat{R}(x)$ as a function of the quantile $x$, with the adapted threshold of $1.04$ \citep[see][]{moinsRhatArxiv}.}
    \label{fig:conv_diag_real_data}
\end{figure}

\begin{figure}
    \centering
    \begin{tabular}{cc}
        Estimation of $(r,\nu,\xi)$ & Estimation of $(r,\nu)$ with $\xi = 0$\\ 
        \includegraphics[trim={0cm 0cm 0cm 0cm},clip,width =0.5\textwidth]{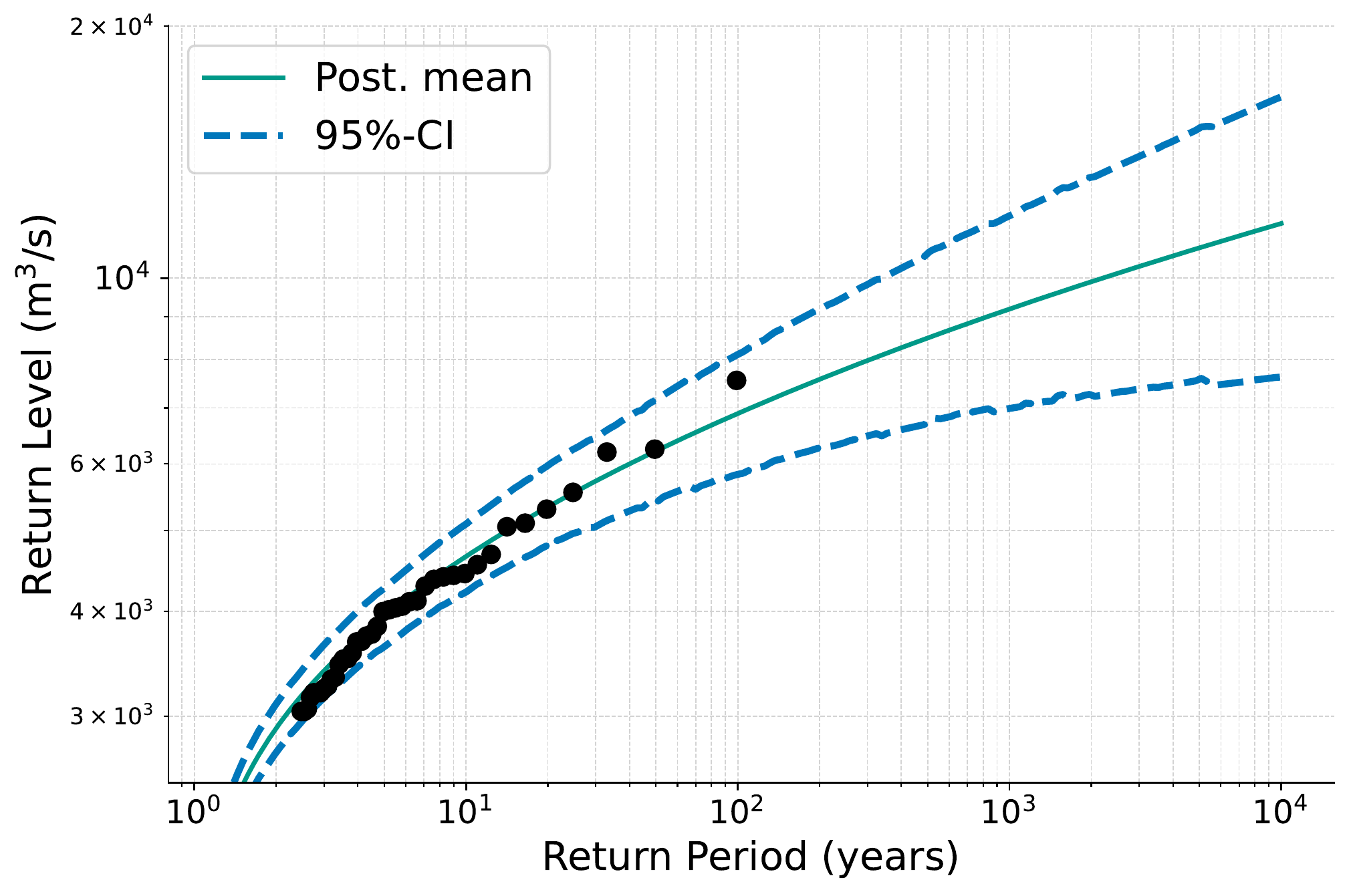}
        &        
        \includegraphics[trim={0cm 0cm 0cm 0cm},clip,width =0.447\textwidth]{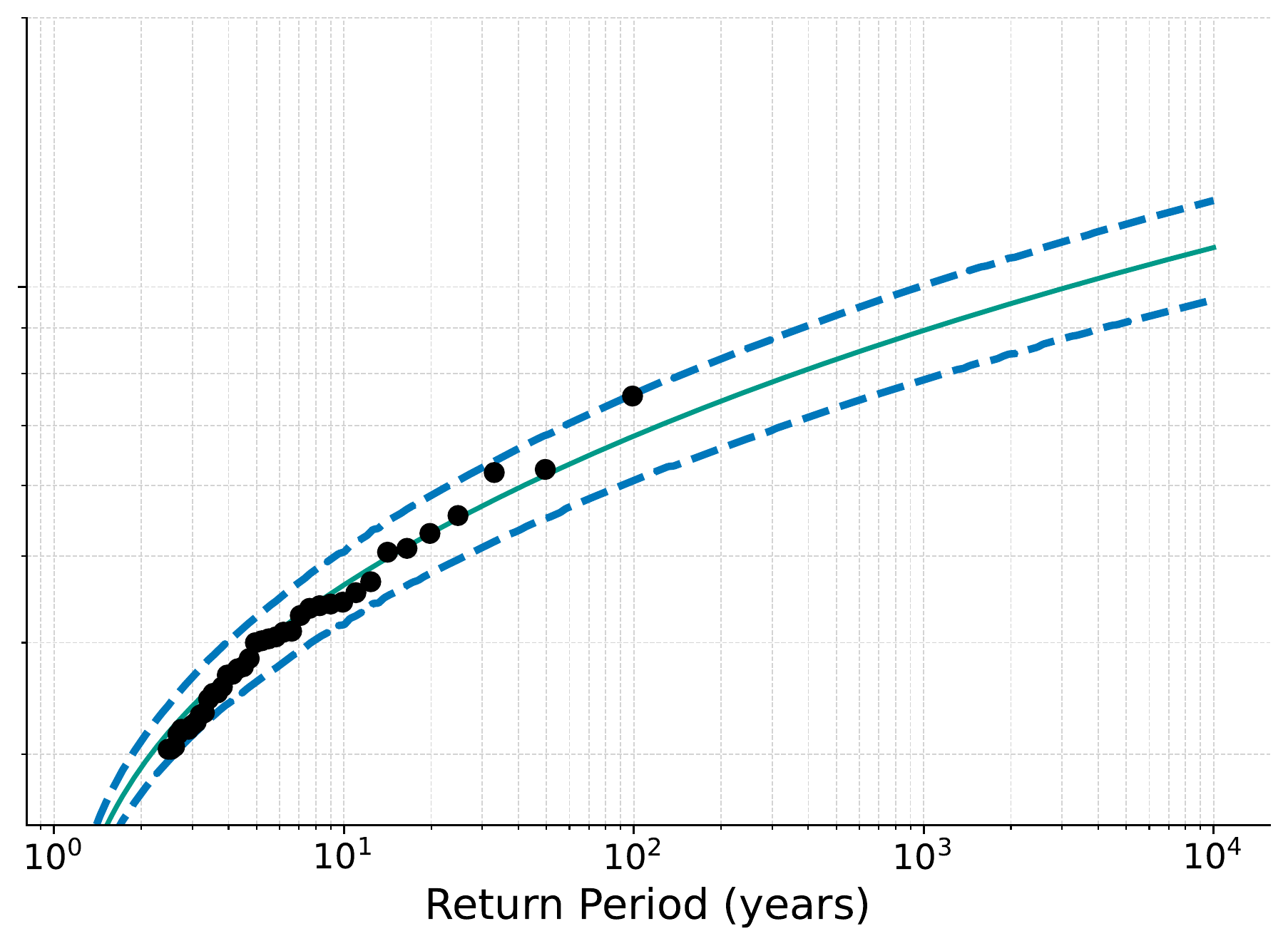}
    \end{tabular}
    \caption{Return levels for annual maxima of Garonne flow data. Full \textcolor{colorblind2}{green} curves correspond to return levels obtained with posterior mean return level, and the dashed ones to the bounds of the $95\%$ credible interval (CI). On the left, all three parameters $(r, \nu, \xi)$ are estimated, while on the right, only $(r, \nu)$ are estimated with the assumption that $\xi = 0$.
    Black points represent the observed annual maxima.}
    \label{fig:return_level}
\end{figure}

\paragraph{Return level estimation}
We are interested in estimating the $T$-year return level $\ell_T$, which is exceeded on average once every $T$ years.
This is obtained by solving the equation $G(\ell_T \mid \mu, \sigma, \xi) = 1-1/T$, with $G$ the GEV cdf defined in Equation~(\ref{eq:GEV-CDF}): 
\begin{equation}
     \ell_T = \mu - \frac{\sigma}{\xi}\left(1 - (-\log (1-1/T))^{-\xi}\right).
     \label{eq:return_level_def}
\end{equation}
Here, as the data span $99$ years, we fix $m=99$ in order to obtain parameters associated with annual maxima.
The same setup as in Section~\ref{subsec:simulations} is then run with $5\,000$ draws from Metropolis--Hastings algorithm with the orthogonal parameterization. 
Convergence diagnostic values are reported in Figure~\ref{fig:conv_diag_real_data} and show no evidence of \cor{lack of} convergence, along with a very satisfactory effective sample size for estimation (final values can be found in Table~\ref{tab:post_summary} along with $\hat{R}_\infty$ for each parameter).
Results of posterior summaries for $(\mu, \sigma, \xi)$ are reported in Table~\ref{tab:post_summary}: 
looking at the posterior for $\xi$, the three maximum domains of attraction cannot be excluded, although the 95\% credible interval (CI) is tight around zero.
This may suggest that $\xi = 0$ and an exponential decrease of the survival function.
Return levels for annual maxima are \cor{displayed} in the left panel of Figure~\ref{fig:return_level}, and show that the model seems to fit the data correctly.
These curves are obtained by computing the mean and $2.5\%$/$97.5\%$ quantiles on the posterior distribution of $\ell_T$ for any given return period $T$. 
This is more accurate than the version where pointwise posterior quantities of $(\mu, \sigma, \xi)$ are plugged in Equation~(\ref{eq:return_level_def}) \citep[see~][for a comparison]{jonathan2021uncertainties}.
The obtained posterior mean of $\ell_T$, is $6\,949$~m$^3$/s for the 100-year level and $9\,266$~m$^3$/s for the $1\,000$-year one.
These results corroborate a study conducted in \cite{albert2020extreme}, where the estimated value of $10\,000$~m$^3$/s for the $1\,000$-year return level belongs to the credible interval in Figure~\ref{fig:return_level}.

\begin{table}
    \centering
    \begin{tabular}{lccccc}
        \toprule
         & Post. Mean & Post. SD & $95\%$-CI &  ESS &  $\hat{R}_\infty$ \\
        \midrule
        $\mu$ & $2\,560.8$ & $84.1$ & [$2\,409.8$, $2\,724.1$] &  $3\,473$ & $\approx 1.0$ \\
        $\sigma$ & $919.6$ & $73.2$ &  [$787.2$, $1\,063.3$] & $2\,709$ &   $\approx 1.0$ \\
        $\xi$ & $0.015$ &  $0.077$ & [$-0.120$, $0.164$] & $2\,702$ &  $\approx 1.0$ \\
        \bottomrule
    \end{tabular}
    \caption{Posterior summaries (mean, standard deviation (SD), credible interval (CI) at 95$\%$) and convergence diagnostics (ESS and $\hat{R}_\infty$) for $(\mu, \sigma, \xi)$ associated with annual maxima $(m=99)$.}
    \label{tab:post_summary}
\end{table}

\paragraph{Prior influence on the return level estimation uncertainty}

\begin{figure}
    \centering
    \begin{tabular}{cc}
        \includegraphics[trim={0cm 0cm 0cm 0cm},clip,width =0.48\textwidth]{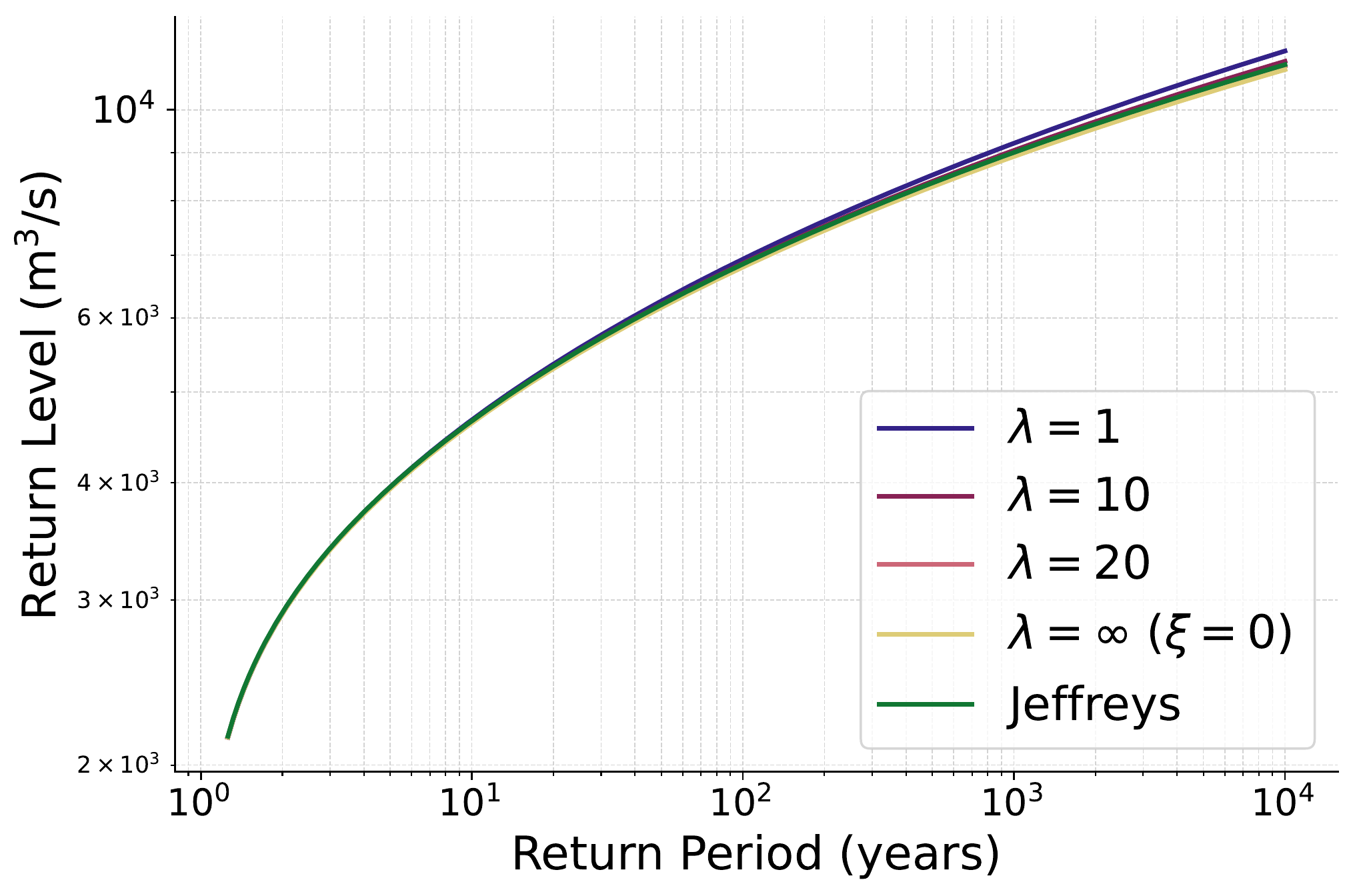}
        &
        \includegraphics[trim={0cm 0cm 0cm 0cm},clip,width =0.48\textwidth]{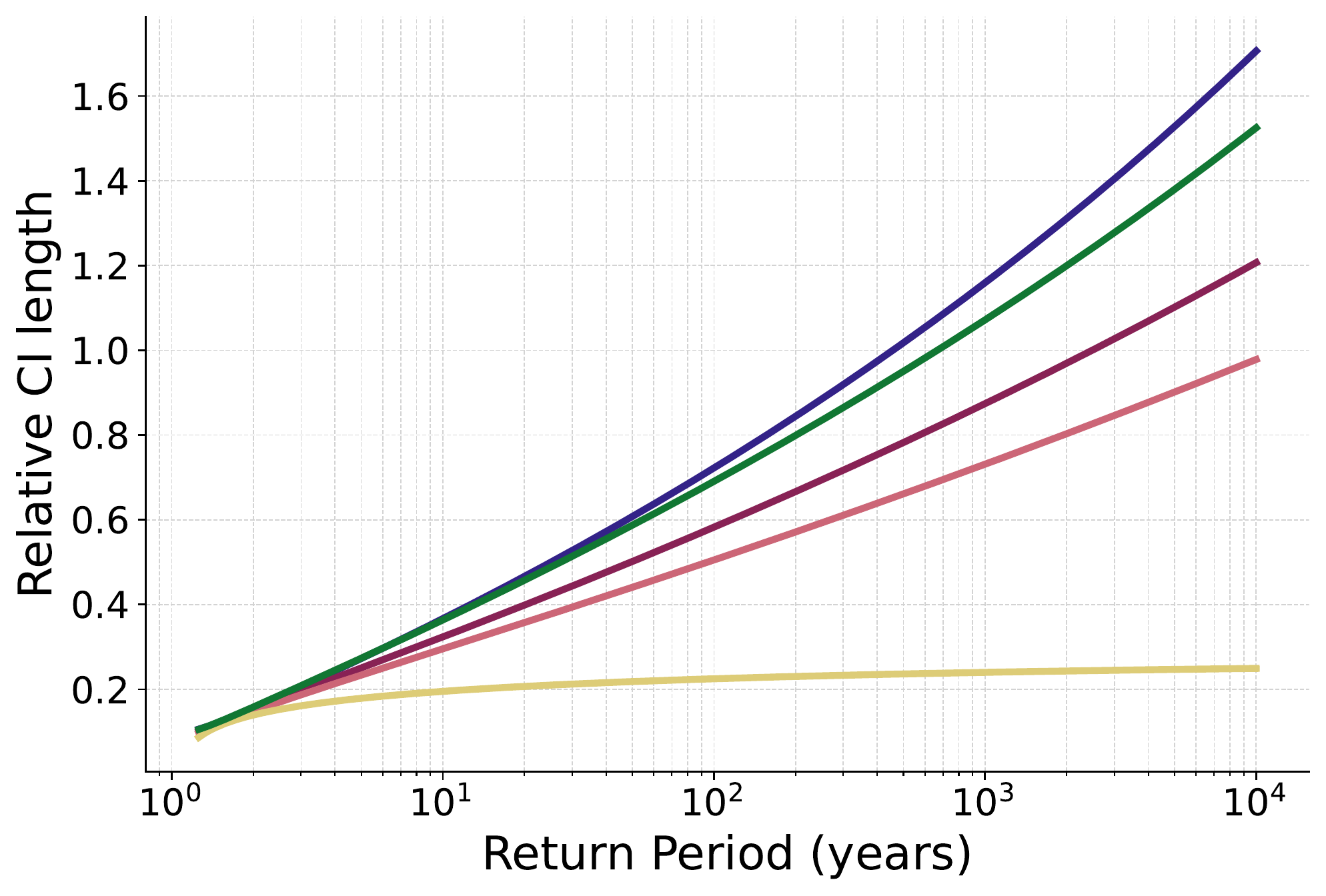}
    \end{tabular}
    \caption{%\textcolor{orange}{[Sur le plot de gauche, je me demande si on ne verrait pas mieux les diff entre les courbes en échelle non log en ordonnée. là on ne distingue rien, pas très utile en soi.]} 
    Comparison of return levels with different priors as functions of return period (log scale). On the left: return levels with posterior mean parameters. On the right: return level credible interval (CI) length relative to the point estimate (in $\%$).}
    \label{fig:relative_hdi}
\end{figure}

Looking at the posterior distribution for $\xi$, one can reasonably make the assumption that $\xi = 0$ and therefore assume an exponential decrease for the survival function of the river flow.
In this case, the remaining location parameter $\mu$ and scale parameter $\sigma$ can be estimated with fixed $\xi = 0$. The resulting posterior summaries are very close to the ones of Table~\ref{tab:post_summary}.
As a result, the return level curves with posterior mean parameters (see Figure~\ref{fig:return_level}) are very similar in both cases.
However, as the uncertainty on the shape parameter is excluded when fixing $\xi = 0$, the return levels credible intervals change drastically and become very concentrated around means, as shown in the right panel of Figure~\ref{fig:return_level}.
In fact, this reflects that most of the uncertainty on the estimated return level is due to the estimation of the shape parameter, and so knowing its value \cor{greatly} facilitates the extrapolation.
PC priors allow \cor{us} to navigate between these two extreme cases thanks to the hyperparameter $\lambda$. 
Looking at the left panel of Figure~\ref{fig:relative_hdi}, it appears that the return level curves associated with posterior means are not affected by those differences of priors.
However, the larger $\lambda$, the more information is added about the closeness of $\xi$ to zero, and the smaller the length of the credible interval (note however that this does not give any guarantee on the estimation bias). 
This behaviour is illustrated on the right panel of Figure~\ref{fig:relative_hdi}: \cor{denoting} by $\ell^{(\mathrm{m})}_T$, $\ell^{(2.5\%)}_T$, and $\ell^{(97.5\%)}_T$ respectively the posterior mean, and the posterior quantiles at $2.5\%$ and $97.5\%$ of the return level, then the right plot in Figure~\ref{fig:relative_hdi} displays the length of the credible interval for the return level estimation, relatively to the estimator $\ell^{(\mathrm{m})}_T$: $(\ell^{(97.5\%)}_T - \ell^{(2.5\%)}_T)/\ell^{(\mathrm{m})}_T$.
This ratio is expected to grow with $T$, as the uncertainty increases in the tail.
When $\lambda = 1$, this growth is similar to the one associated with Jeffreys prior, which can be seen as a noninformative case.
For example, \corr{one} can see that the size of the credible interval is already greater than the posterior estimation for the $1\,000$-year return level (ratio greater than one).  
Using $\lambda = 10$ corresponds to a confidence of $95\%$ of having $\xi$ between $-0.3$ and $0.3$ with the version approximated by a Laplace distribution (see the table in Figure~\ref{fig:pc_prior}), reduces by approximately $20\%$ the size of the credible interval for $T=1\,000$.
The length when $\xi$ is fixed at zero is drastically lower than in the other cases, even those concerning PC priors with large $\lambda$ values.
%In our case where this dataset has already been studied in the past and a value very close to zero was expected, a choice of $\lambda = 10$ seems reasonable for the PC prior.

\section{Conclusion}

In this paper we demonstrate the benefits of using an orthogonal parameterization in the sense of \cite{Jeffreys61} for Bayesian inference of extreme value models.
First, orthogonal parameters facilitate the convergence of MCMC algorithms such as Metropolis--Hastings or NUTS (Section~\ref{subsec:sharkey} and \ref{sec:sharkey}). 
This improvement is ``free'' in the sense that it is obtained at no extra computational cost, except a simple change of variable if one interest lies in the original parameters $(\mu, \sigma, \xi)$.
This conclusion is confirmed by convergence diagnostics such as autocorrelation, effective sample size, and local $\hat R$, on simulations in the three maximum domains of attraction (Section~\ref{subsec:simulations} and \ref{sec:additional_exp}).

Secondly, the orthogonal parameterization also facilitates the computation of Jeffreys prior (Section~\ref{subsec:jeffreys}): we show that this uninformative prior is defined for $\xi > -1/2$ and is improper, but leads to a proper posterior.
Posterior propriety is a necessary condition for using this prior in practice when no external information is available.
However, this uninformative case is actually far from the reality of most of the applications: even without any expert information, a shape parameter in the range $(-1,1)$ already includes a vast majority of the distributions arising in natural phenomena. 
Therefore as an alternative, a PC prior on $\xi$ can be used instead and allows \cor{users} to control the prior knowledge \cor{they} want to include on $\xi$ (Section~\ref{subsec:pc_prior}). 
In particular, it penalizes the values of $\xi$ that move away from 0, and navigate between the uninformative case and the deterministic one where $\xi = 0$.
In addition to its flexibility, this prior enjoys the same advantages as Jeffreys prior: invariance to reparameterization and posterior propriety.
Additionally, it can be defined without any restriction for $\xi$ if one uses the approximation by a Laplace distribution (otherwise, $\xi < 1$).
This prior information on $\xi$ impacts the posterior uncertainty around the return level estimation.
By applying our framework on river flow data (Section~\ref{subsec:real_data}), we showed that the length of the credible interval for the return level can be significantly reduced by adding prior information of $\xi$.
However, the uncertainty around the return level can be quantified differently, by using the quantiles of the posterior predictive distribution defined in~(\ref{eq:post_pred_def}), see \cite{fawcett2018bayesian} for a comparison. 
In future work, it would be interesting to also \cor{investigate} the influence of the prior on the posterior predictive return levels.

\section*{Acknowledgement}

We would like to thank the anonymous reviewers and an Editor for their careful reading and for providing us with valuable comments that helped us improving the manuscript. S.~Girard acknowledges the support of the Chair Stress Test, Risk Management and Financial Steering, led by the  \'Ecole polytechnique and its Foundation and sponsored by BNP Paribas.
J. Arbel acknowledges the support of the French National Research Agency (ANR-21-JSTM-0001).

\bibliography{references}

\appendix
% \begin{appendices}

% \renewcommand\thefigure{\thesection.\arabic{figure}}
% \renewcommand\thetable{\thesection.\arabic{table}}
\stepcounter{appsection}

\section{Approaching orthogonality by choosing $m=n_u$}
\label{sec:sharkey}

\cite{Sharkey2017} \cor{suggests} to take a value of the scaling factor $m$ that minimises the off-diagonal terms of the asymptotic covariance matrix (that is the inverse Fisher information matrix), denoted by $\text{ACov}:= \mathcal{I}^{-1}(\mu, \sigma, \xi)$.
Those terms exist only if $\xi > -1/2$ (see Proposition~\ref{prop:jeffreys_proper} and its proof in \ref{sec:proof_prior}) and can be written as functions of $x=-\frac{1}{\xi}\log\left\{1+\xi\left(\frac{u-\mu}{\sigma}\right)\right\}_+$, $\sigma$, and $\xi$ as:
\begin{align*}
    \text{ACov}_{\mu, \sigma} &= 
    \frac{\sigma^2}{m\xi^2} e^x\left(\xi^3 + (1+\xi)(1+2\xi + \xi(1+\xi)x^2 - (1+3\xi)x + e^{-\xi x}(1+2\xi)(x-1))\right),\\
    \text{ACov}_{\mu, \xi} &= \frac{\sigma}{m\xi^2}e^x(1+\xi) \left(\xi(1+\xi)x - (1+2\xi)(1-e^{-\xi x})\right), \\
    \text{ACov}_{\sigma, \xi} &= \frac{\sigma}{m}e^x(1+\xi)\left((1+\xi)x - 1\right).
\end{align*}
Denote by $\rho_{\cdot,\cdot}$ the asymptotic correlation between two out of the three parameters, the authors note that a range of values may also work for $m$ between $m_1$ and $m_2$, where 
$$
m_1 = \underset{m}{\operatorname{argmin}} \{ |\rho_{\mu, \sigma}| + |\rho_{\mu, \xi}|\} \mbox{ and } m_2 = \underset{m}{\operatorname{argmin}} \{ |\rho_{\mu, \sigma}| + |\rho_{\sigma, \xi}|\}.
$$
They also find on their experiments that $m_1$ cancels $\rho_{\mu, \sigma}$, and that $m_2$ cancels $\rho_{\sigma, \xi}$.
A numerical method is used in \cite{Sharkey2017} to approximate $m_1$ and $m_2$ as functions of $\xi$.
Therefore, this approach requires to study the roots $x_1$ of $\text{ACov}_{\sigma, \xi}$ and  $x_2$ of $\text{ACov}_{\mu, \sigma}$ to respectively \cor{derive} $\hat{m}_1(\xi)$ and $\hat{m}_2(\xi)$. 
Without any approximation, we directly have $x_1 = {1}/(1+\xi)$ as the unique root for $\text{ACov}_{\sigma, \xi}$.
Moreover, as $\xi > -{1}/{2}$, we have $x_1 > 0$, which motivates us to study the sign of the root $x_2$ for $\text{ACov}_{\mu, \sigma}$. 
Indeed, if $x_2$ is unique and $x_2 < 0$, then the choice $x=0$ which cancels the third asymptotic covariance $\text{ACov}_{\mu, \xi}$  will always be reasonable as it will stay in the targeted interval, between the two other roots.
In addition, $x=0$ corresponds to the choice $m=r$ (which in practice translates into $m=n_u$), and is a simple choice as it does not require any estimation of $\xi$.
The interest of the choice $m=n_u$ has already been mentioned in  \cite{Wadsworth2010} to improve the mixing property of the chain.
Unfortunately, a study of function $x \mapsto \text{ACov}_{\mu, \sigma}(x)$ shows that the properties of uniqueness and positivity of $x_2$ are only valid in the case where $\xi > 0$. 
In that case, \cor{the works} of \cite{Wadsworth2010} and \cite{Sharkey2017} corroborate the choice of $m=n_u$.
However, it is not the case anymore when $-{1}/{2} < \xi < 0$.
It can be shown that $x_2$ is not negative here, and worse, may not be unique.
This can be seen as a counter-indication for frameworks that aim at reducing the three asymptotic covariances at the same time by tuning the scaling factor $m$.

\stepcounter{appsection}

\section{Proofs}
\label{sec:proof_prior}

\paragraph{Proof of Proposition~\ref{first-prop}}
The log-likelihood $l$ using the $(r, \nu, \xi)$ parameterization of Equation~(\ref{eq:reparam}) can be written as:
\begin{multline*}
    l(r, \nu, \xi \mid \boldsymbol{x}, n_u) = -r + n_u \log\left(\frac{r}{m}\right) - n_u \log(\nu) + n_u \log(1+\xi) \\
    - \left(1+\frac{1}{\xi}\right)\sum_{i=1}^{n_u} \log\left\{1+ \frac{\xi(1+\xi)}{\nu}(x_i-u) \right\}_+.
\end{multline*}
Under this form, \corr{one} can directly see that $r$ is orthogonal to $\nu$ and $\xi$. 
The second derivatives of $l$ are
\begin{align*}
    \frac{\partial^2 l}{\partial r^2} &= -\frac{n_u}{r^2},\qquad
    \frac{\partial^2 l}{\partial r \partial \nu} =  0,\qquad
    \frac{\partial^2 l}{\partial r \partial \xi} =  0,\\
    \frac{\partial^2 l}{\partial \nu^2} &= 
        \frac{n_u}{\nu^2} 
        + \frac{\xi(1+\xi)^3}{\nu^4}\sum_{i=1}^{n_u} \frac{(x_i- u)^2}{\left\{1+\frac{\xi(1+\xi)}{\nu}(x_i-u)\right\}_+^2}
        - \frac{2(1+\xi)^2}{\nu^3}\sum_{i=1}^{n_u} \frac{(x_i- u)}{\left\{1+\frac{\xi(1+\xi)}{\nu}(x_i-u)\right\}_+},\\
    \frac{\partial^2 l}{\partial \nu \partial \xi} &= 
        \cor{-}\frac{(1+2\xi)(1+\xi)^2}{\nu^3} \sum_{i=1}^{n_u} \frac{(x_i- u)^2}{\left\{1+\frac{\xi(1+\xi)}{\nu}(x_i-u)\right\}_+^2}
        \cor{+} \frac{2(1+\xi)}{\nu^2} \sum_{i=1}^{n_u} \frac{(x_i- u)}{\left\{1+\frac{\xi(1+\xi)}{\nu}(x_i-u)\right\}_+},\\
    \frac{\partial^2 l}{\partial \xi^2} &= 
        -\frac{{n_u}}{(1+\xi)^2} 
        + \frac{(1+2\xi)^2(1+\xi)}{\xi \nu^2}\sum_{i=1}^{n_u} \frac{(x_i- u)^2}{\left\{1+\frac{\xi(1+\xi)}{\nu}(x_i-u)\right\}_+^2}\\
        &+\frac{2(1+\xi-\xi^2)}{\xi^2 \nu}\sum_{i=1}^{n_u} \frac{(x_i- u)}{\left\{1+\frac{\xi(1+\xi)}{\nu}(x_i-u)\right\}_+}
        -\frac{2}{\xi^3}\sum_{i=1}^{n_u} \log\left\{1+\frac{\xi(1+\xi)}{\nu}(x_i- u)\right\}_+.
\end{align*}
\cor{Focussing on} the expectations, as we observe a Poisson process, the information is contained in the number $n_u$ of observed points (we write $N_u$ the corresponding random variable) and the position of jumping events $x_i$ (we write $X_i$ the corresponding random variable, with the same distribution as $X$).
Here, $N_u$ is distributed according to a Poisson distribution with parameter $r$, and $X-u$ is a GPD random variable with parameters $(\frac{\nu}{1+\xi}, \xi)$.
For example, deriving the following expectations is the cornerstone to obtain the Fisher information matrix:
\begin{align*}
    \mathbb{E}_{{N_u},X}&\left[\sum_{i=1}^{N_u} \frac{(X_i - u)^2}{\left\{1+\frac{\xi(1+\xi)}{\nu}(X_i-u)\right\}_+^2}\right]    
    = \mathbb{E}_{{N_u}}\left[\mathbb{E}_{X \mid {N_u}}\left[\sum_{i=1}^{N_u} \frac{(X_i - u)^2}{\left\{1+\frac{\xi(1+\xi)}{\nu}(X_i-u)\right\}_+^2}\right]\right] \\
    &= \mathbb{E}_{{N_u}}\left[N_u\right] \mathbb{E}_{X\mid {N_u}}\left[\frac{(X - u)^2}{\left\{1+\frac{\xi(1+\xi)}{\nu}(X-u)\right\}_+^2}\right]\\
    &= r \frac{1+\xi}{\nu}\int_u^{+\infty} (x - u)^2\left\{1+\frac{\xi(1+\xi)}{\nu}(x-u)\right\}_+^{-\frac{1}{\xi} -3} \dd x.
\end{align*} %By distinguishing two cases according to the sign of $\xi$, we obtain that t
The above integral exists provided $\xi > -1/2$ and we obtain
\begin{equation*}
    \mathbb{E}_{{N_u},X}\left[\sum_{i=1}^{N_u} \frac{(X_i - u)^2}{\left\{1+\frac{\xi(1+\xi)}{\nu}(X_i-u)\right\}_+^2}\right]
    = \frac{2r\nu^2}{(1+\xi)^3(1+2\xi)}.
\end{equation*}
Similarly, the remaining expected values can be written as
\begin{align*}
    \mathbb{E}_{{N_u},X}\left[\sum_{i=1}^{N_u} \frac{(X_i - u)}{\left(1+\frac{\xi(1+\xi)}{\nu}(X_i-u)\right)}\right] 
    &= \frac{r\nu}{(1+\xi)^2},\\
    \mathbb{E}_{{N_u},X}\left[\sum_{i=1}^{N_u} \log \left(1+\frac{\xi(1+\xi)}{\nu}(X_i-u)\right)\right] 
    &= r \xi.
\end{align*}
Plugging these values into the Fisher coefficients yields the result:
$$
I(r, \nu, \xi) = \text{diag}\left(\frac{1}{r},  \frac{r}{\nu^2(1+2\xi)}, \frac{r}{(1+\xi)^2}\right).
$$

\paragraph{Proof of Proposition~\ref{prop:jeffreys_proper}}
Let us show that the following integral exists for any $n_u\geq 1$:
\begin{equation*}
    C_{n_u}
    = 
    \int_{\mathcal{S}}
    \frac{r^{1/2}e^{-r}}{\nu (1+\xi) (1 + 2 \xi)^{1/2}}
    \left(\frac{r(1+\xi)}{m\nu}\right)^{n_u}
        \prod_{i=1}^{n_u} \left(1+ \frac{\xi(1+\xi)}{\nu}(x_i-u) \right)^{-1-\frac{1}{\xi}}
        \dd r \dd\nu \dd\xi,
\end{equation*}
where $\mathcal{S}$ is the integration domain:
\begin{equation*}
    \mathcal{S} = \left\{ (r, \nu, \xi) \in \mathbb{R}^3 \text{ s.t. } 
        \xi > -\frac{1}{2},\,
        r > 0,\,
        \nu \geq \{-\xi(1+\xi)((\max_i x_i)-u)\}_+
    % \left\{
    % \begin{array}{l}
    %     \xi > -\frac{1}{2}\\
    %     r > 0\\
    %     \nu \geq \left\{-\xi(1+\xi)((\max_i x_i)-u)\right\}_+
    % \end{array}
    % \right.
    \right\}.
\end{equation*}
Let us consider the case of one observation ($n_u=1$):
\begin{align*}
    &\int_{-\frac{1}{2}}^{+\infty}
    (1+2\xi)^{-\frac{1}{2}}
    \int_{0}^{+\infty}
    r^{\frac{3}{2}} e^{-r}
    \int_{\{-\xi(1+\xi)(x-u)\}_+}^{+\infty} 
    \nu^{-2}
    \left(1+ \frac{\xi(1+\xi)}{\nu}(x-u) \right)^{-\frac{1}{\xi} -1} \dd\nu \dd r \dd\xi\\
    &=\int_{-\frac{1}{2}}^{0}
    (1+2\xi)^{-\frac{1}{2}}
    \int_{0}^{+\infty}
    r^{\frac{3}{2}} e^{-r}
    \int_{-\xi(1+\xi)(x-u)}^{+\infty} 
    \nu^{-2}
    \left(1+ \frac{\xi(1+\xi)}{\nu}(x-u) \right)^{-\frac{1}{\xi} -1} \dd\nu \dd r \dd\xi\\
    &+\int_{0}^{+\infty}
    (1+2\xi)^{-\frac{1}{2}}
    \int_{0}^{+\infty}
    r^{\frac{3}{2}} e^{-r}
    \int_{0}^{+\infty} 
    \nu^{-2}
    \left(1+ \frac{\xi(1+\xi)}{\nu}(x-u) \right)^{-\frac{1}{\xi} -1} \dd\nu \dd r \dd\xi\\
    &=\int_{-\frac{1}{2}}^{0}
    (1+2\xi)^{-\frac{1}{2}}
    \int_{0}^{+\infty}
    r^{\frac{3}{2}} e^{-r}
    \left[\frac{1}{(1+\xi)(x-u)}\left(1+ \frac{\xi(1+\xi)}{\nu}(x-u) \right)^{-\frac{1}{\xi}}\right]_{-\xi(x-u)\left(\frac{r}{m}\right)^\xi}^{+\infty} \dd r \dd\xi\\
    &+\int_{0}^{+\infty}
    (1+2\xi)^{-\frac{1}{2}}
    \int_{0}^{+\infty}
    r^{\frac{3}{2}} e^{-r}
    \left[\frac{1}{(1+\xi)(x-u)}\left(1+ \frac{\xi(1+\xi)}{\nu}(x-u) \right)^{-\frac{1}{\xi}}\right]_{0}^{+\infty} \dd r \dd\xi\\
    &= \frac{1}{(x-u)}\int_{-\frac{1}{2}}^{+\infty}
    (1+\xi)^{-1}(1+2\xi)^{-\frac{1}{2}}
    \int_{0}^{+\infty}
    r^{\frac{3}{2}} e^{-r} \dd r \dd\xi\\
    &= \frac{3\pi^{\frac{3}{2}}}{4(x-u)} < \infty.
\end{align*}
Therefore, the posterior is proper for $n_u=1$. It is well-known that it stays so for $n_u > 1$ as can be seen  by induction. For instance for $n_u=2$, the posterior writes
\begin{align*}
    p(\boldsymbol{\theta} \mid x_1, x_2) 
    \propto p(x_1, x_2 \mid \boldsymbol{\theta})p(\boldsymbol{\theta})
    = p(x_2 \mid \boldsymbol{\theta})p(x_1 \mid \boldsymbol{\theta})p(\boldsymbol{\theta})
    \propto p(x_2 \mid \boldsymbol{\theta})p(\boldsymbol{\theta}\mid x_1) \leq p(\boldsymbol{\theta}\mid x_1)
\end{align*}
which is integrable.

\paragraph{Proof of Proposition~\ref{prop:pc_proper}}

Similarly to the proof of Proposition~\ref{prop:jeffreys_proper}, the aim is to show the existence of the following integral for any $n_u$:
\begin{equation*}
    C_{n_u} 
    = \int_{\mathcal{S}}
    \frac{p_{\text{PC}}(\xi \mid \lambda)}{\nu}
    e^{-r}\left(\frac{r}{m}\right)^{n_u} \left(\frac{\nu}{1+\xi}\right)^{-{n_u}}
        \prod_{i=1}^{n_u} \left(1+ \frac{\xi(1+\xi)}{\nu}(x_i-u) \right)^{-1-\frac{1}{\xi}}
        \dd r \dd\nu \dd\xi,
\end{equation*}
with $p_{\text{PC}}(\xi \mid \lambda)$ defined in Equation~(\ref{eq:pc_prior}), and $\mathcal{S}$ the following integration domain: 
\begin{equation*}
    \mathcal{S} = \left\{ (r, \nu, \xi) \in \mathbb{R}^3 \text{ s.t. } 
        \xi < 1,\,
        r > 0,\,
        \nu \geq \{-\xi(1+\xi)((\max_i x_i)-u)\}_+
    \right\}.
\end{equation*}
In the general case for $n_u$, we have
\begin{align*}
    C_{n_u} % &= 
    % \int_{-\infty}^1\int_{\{-\xi(1+\xi)(x-u)\}_+}^{+\infty} \frac{p_{\text{PC}}(\xi \mid \lambda)}{\nu}\left(\frac{\nu}{1+\xi}\right)^{-{n_u}}
    % \prod_{i=1}^{n_u} \left(1+ \frac{\xi(1+\xi)}{\nu}(x_i-u) \right)^{-1-\frac{1}{\xi}} \int_{0}^{+\infty} \left(\frac{r}{m}\right)^{n_u} e^{-r} \dd r \dd\nu \dd\xi\\
    &=
    \frac{\Gamma({n_u}+1)}{m^{n_u}} \int_{-\infty}^1\int_{\{-\xi(1+\xi)(x-u)\}_+}^{+\infty} \frac{p_{\text{PC}}(\xi \mid \lambda)}{\nu}\left(\frac{\nu}{1+\xi}\right)^{-{n_u}}\\
    &\qquad\qquad\qquad\qquad\qquad\qquad\qquad\qquad\prod_{i=1}^{n_u} \left(1+ \frac{\xi(1+\xi)}{\nu}(x_i-u) \right)^{-1-\frac{1}{\xi}} \dd\nu \dd\xi \\
    &= \frac{\Gamma(n_u+1)}{m^{n_u}} \int_{-\infty}^1\int_{\{-\xi(x-u)\}_+}^{+\infty} \frac{p_{\text{PC}}(\xi \mid \lambda)}{\sigma}\sigma^{-{n_u}}
    \prod_{i=1}^{n_u} \left(1+ \xi\left(\frac{x_i-u}{\sigma}\right) \right)^{-1-\frac{1}{\xi}} \dd\sigma \dd\xi.
\end{align*}
The remaining integral corresponds to the normalizing constant of the posterior distribution of a GPD model with a prior of the form $p(\sigma, \xi) \propto p(\xi)/\sigma$. 
Since $p(\xi)$ is a proper density, Theorem 1 in \cite{Northrop2016} \cor{allows us to conclude that $C_{n_u}$ is finite} for any $n_u \geq 1$.
Note that this result remains true with $p_{\text{PC}}(\xi \mid \lambda)$ replaced by a Laplace distribution as suggested in Section~\ref{subsec:pc_prior}, \cor{since} the prior on $\xi$ remains proper.

\stepcounter{appsection}

\section{Additional experiments}
\label{sec:additional_exp}

\subsection{Simulations using an Hamiltonian Monte Carlo algorithm}
\label{subsec:NUTS}

\begin{figure}[h]
    \centering
    \includegraphics[trim={0cm 0cm 0cm 0cm},clip,width =\textwidth]{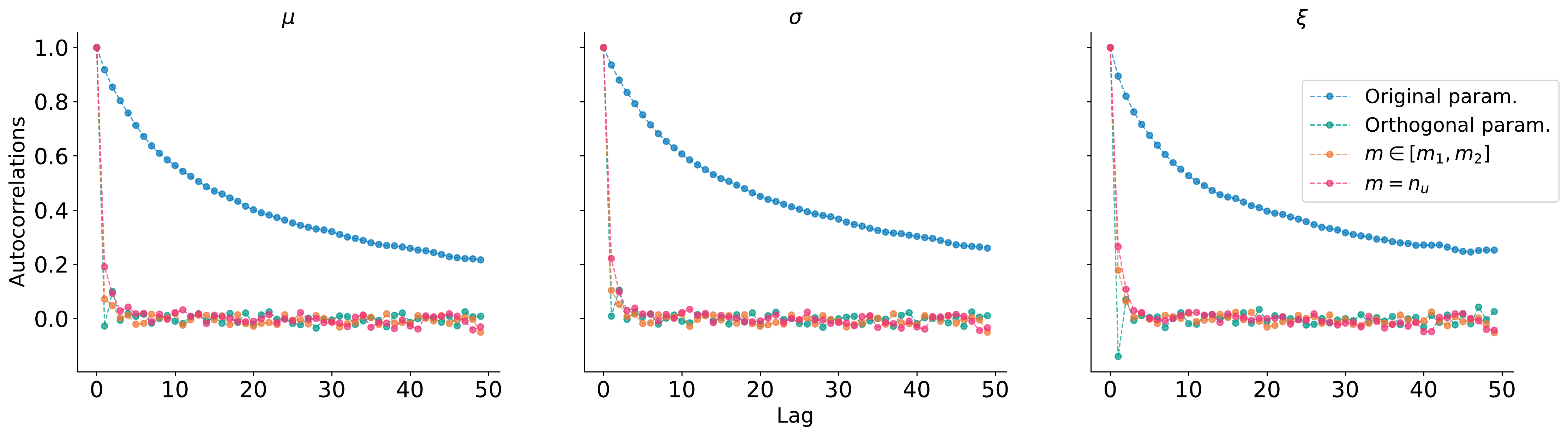}

    \includegraphics[trim={0cm 0cm 0cm 0cm},clip,width =\textwidth]{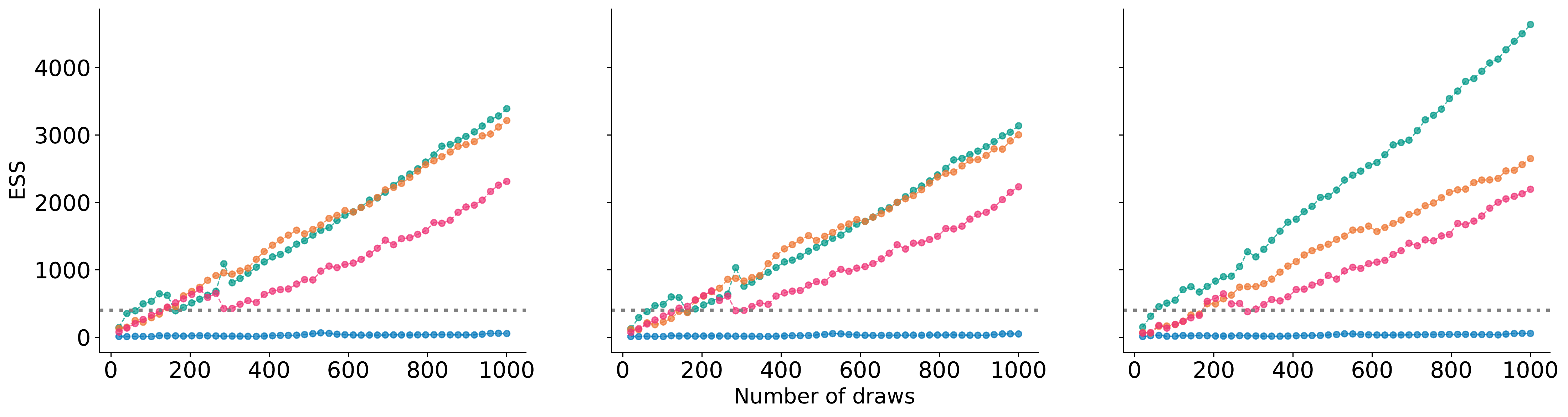}

    \includegraphics[trim={0cm 0cm 0cm 0cm},clip,width =\textwidth]{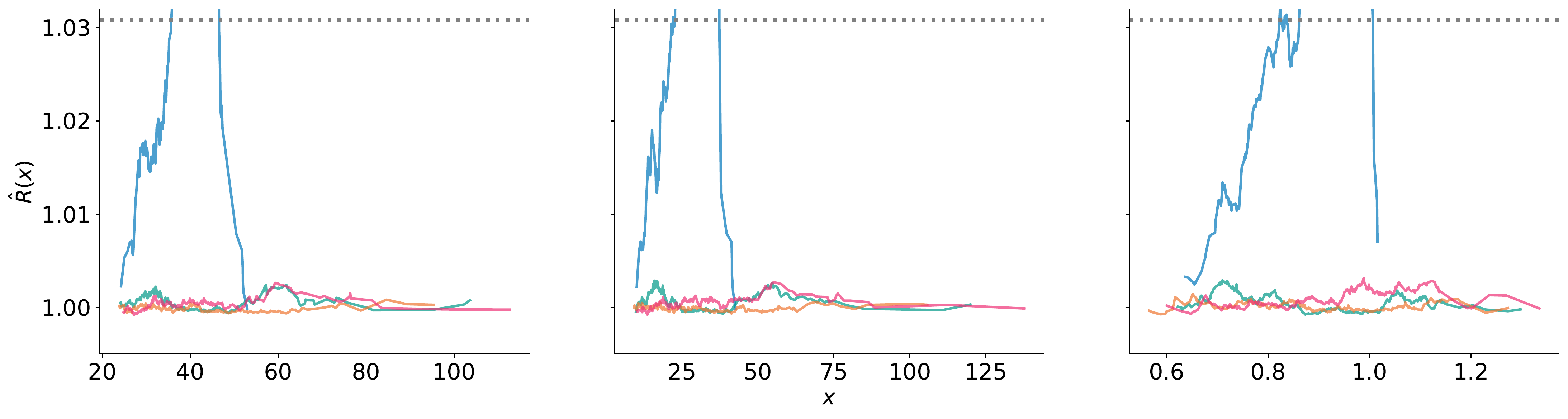}
    \caption{Convergence diagnostic plots for Poisson parameters $(\mu, \sigma, \xi)$ with $\xi > 0$, after $500$ NUTS draws and a burn-in of $1\,000$, for four different parameterizations: the original one (in \textcolor{colorblind1}{blue}), the \cite{Sharkey2017} update with $m \in [\hat{m}_1, \hat{m}_2]$ (in \textcolor{colorblind3}{orange}), the \cite{Wadsworth2010} update with $m = n_u$ (in \textcolor{colorblind4}{magenta}), and the orthogonal parameterization (in \textcolor{colorblind2}{green}).
    Top row: autocorrelations as functions of the lag. 
    Second row: evolution of ESS with the number of draws (the gray line corresponds to value of $400$ recommended in \cite{gelman2013bayesian}).
    Bottom row: $\hat{R}(x)$ as a function of the quantile $x$, with the adapted threshold of $1.03$ \citep[see][]{moinsRhatArxiv}.
    The red curve is truncated for visibility purposes, as it is taking much larger values than the threshold.}
    \label{fig:NUTS_xi_pos}
\end{figure}

Hamiltonian Monte Carlo \citep[HMC,][]{neal2011mcmc} and its variants such as NUTS \citep{hoffman2014no} are MCMC methods with a Markov kernel based on trajectories of particles computed using Hamiltonian dynamics. 
\cor{As a consequence}, the performance of these methods is also sensitive to the choice of the parameterization (see \citealp{Betancourt2019} for a formalization of the problem).
We performed the same experiments as those in Section~\ref{subsec:simulations} and \ref{subsec:simul_appendix}, using 500 NUTS iterations instead of $1\,000$ Metropolis--Hastings draws.
The results obtained here are similar, and show that the orthogonal parameterization improves the efficiency of NUTS sampling.
\cor{The case $\xi>0$  is illustrated in} Figure~\ref{fig:NUTS_xi_pos}  with the same configuration as the one described in the first paragraph of \ref{subsec:simul_appendix}.
We observe similar trends as those in Figure~\ref{fig:PP_xi_positive}: changing the value of $m$ improves convergence, and using the orthogonal parameterization is even better.
Moreover, NUTS seems to be more efficient on the three cases than with Metropolis--Hastings, as the chains seem to be less correlated compared to their equivalent in Figure~\ref{fig:PP_xi_positive}, and the ESS can even be greater than the number of draws. 
% Nevertheless, although the computation time is fast on all methods (a few seconds), it should be noted that it is a bit longer to run 500 NUTS samples than 2000 Metropolis.

\subsection{Simulations in other maximum domains of attraction}
\label{subsec:simul_appendix}

We study the influence of parameterizations for MCMC convergence in cases where $\xi > 0$ and $\xi = 0$.

\paragraph{Example with $\xi > 0$}

\begin{figure}
    % \centering
    % \textbf{Example~\hyperref[fig:PP_xi_positive]{1}:} Poisson process with $\xi > 0$\\
    \includegraphics[trim={0cm 0cm 0cm 0cm},clip,width =1\textwidth]{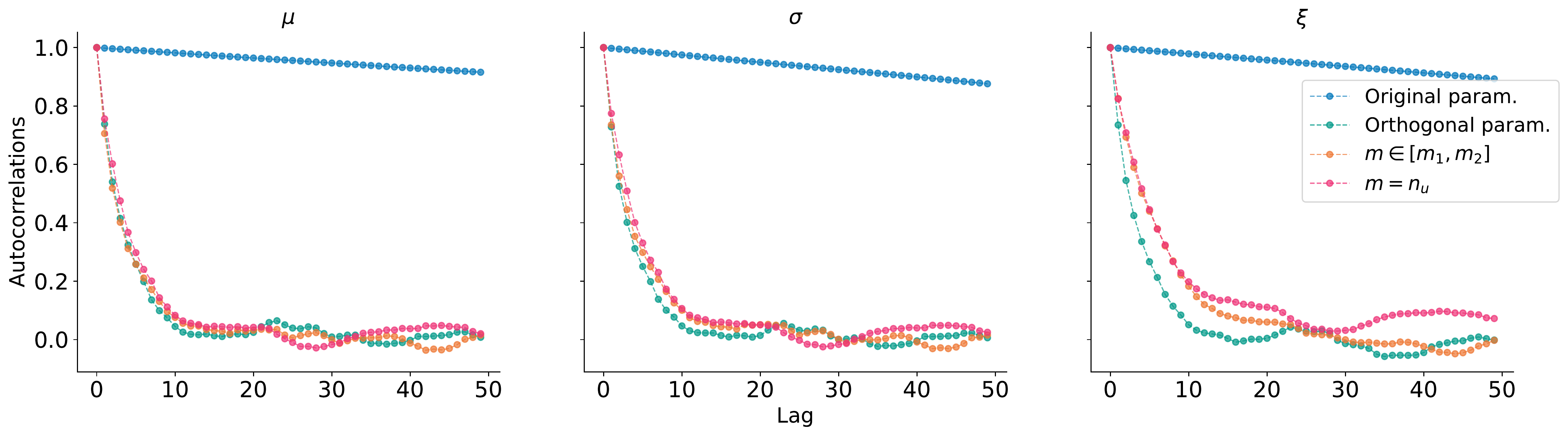}

    \includegraphics[trim={0cm 0cm 0cm 0cm},clip,width =1\textwidth]{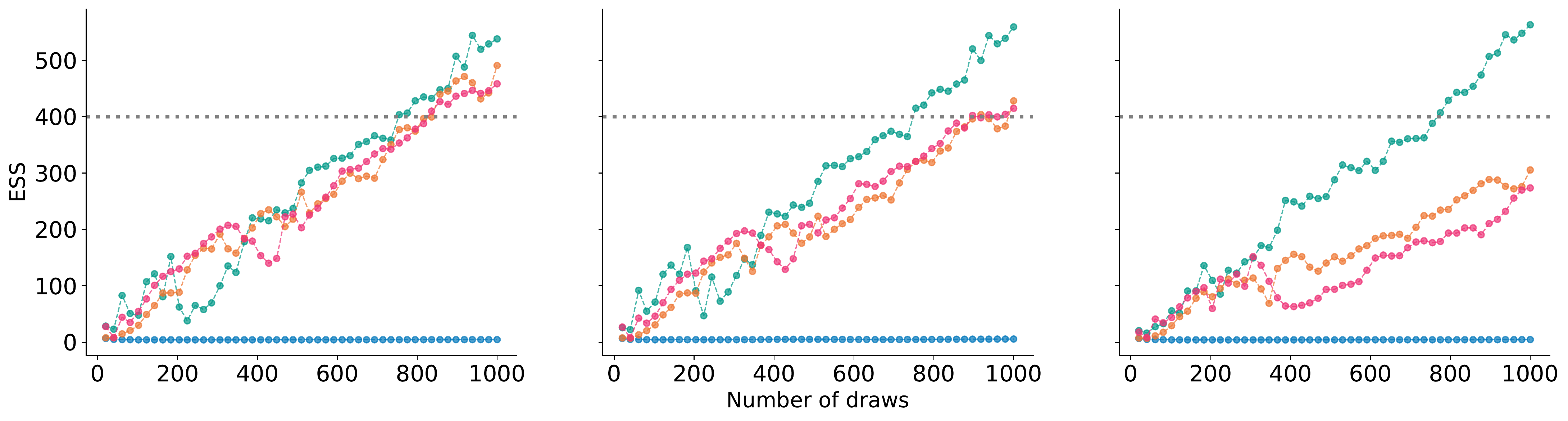}

    \includegraphics[trim={0cm 0cm 0cm 0cm},clip,width =1\textwidth]{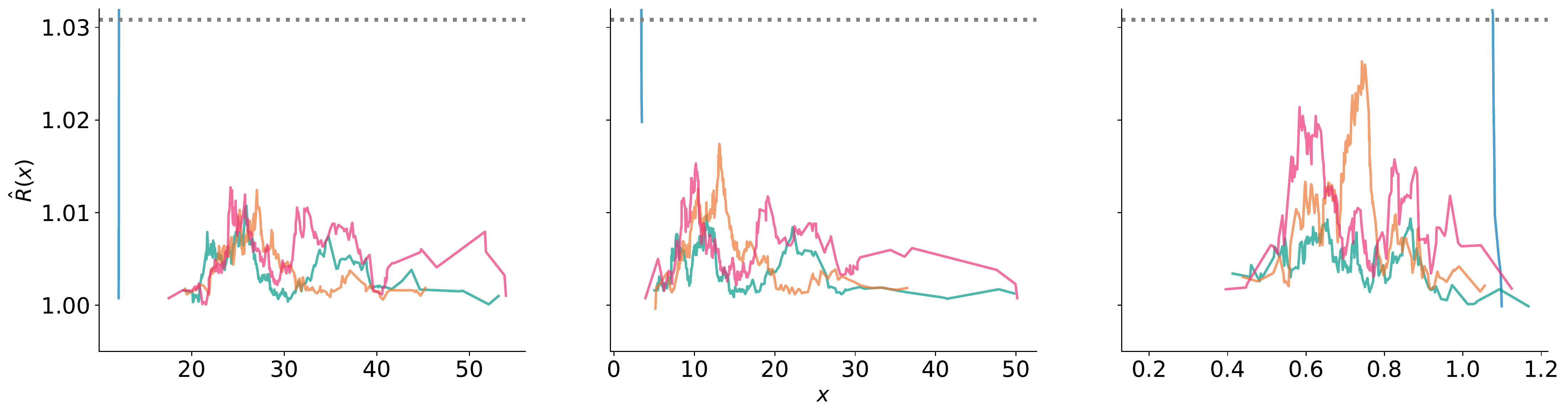}
    \caption{Convergence diagnostic plots for Poisson parameters $(\mu, \sigma, \xi)$ with $\xi > 0$, after $1~000$ Metropolis--Hastings draws and a burn-in of $1\,000$, for four different parameterizations: the original one (in \textcolor{colorblind1}{blue}), the \cite{Sharkey2017} update with $m \in [\hat{m}_1, \hat{m}_2]$ (in \textcolor{colorblind3}{orange}), the \cite{Wadsworth2010} update with $m = n_u$ (in \textcolor{colorblind4}{magenta}), and the orthogonal parameterization (in \textcolor{colorblind2}{green}).
    Top row: autocorrelations as functions of the lag. 
    Second row: evolution of ESS with the number of draws (the gray line corresponds to value of $400$ recommended in \cite{gelman2013bayesian}).
    Bottom row: $\hat{R}(x)$ as a function of the quantile $x$, with the adapted threshold of $1.03$ \citep[see][]{moinsRhatArxiv}.
    The red curve is truncated for visibility purposes, as it is taking much higher values than the  threshold.}
    \label{fig:PP_xi_positive}
\end{figure}

Here, we set $(m, u, \mu, \sigma, \xi) = (5, 10, 30, 15, 0.7)$, which leads to an expected number of observations
$r \approx 239$.
Looking at autocorrelations, ESS and $\hat{R}(x)$ curves in Figure~\ref{fig:PP_xi_positive}, 
we can first confirm the result of \cite{Sharkey2017} about the inefficiency of Metropolis--Hastings \cor{with} the original parameterization: high autocorrelations, high $\hat R(x)$ (around $1.7$ for the highest) and almost zero ESS even after $1\,000$ iterations indicate a severe convergence issue.
Changing the value of $m$ before the MCMC algorithm as suggested by \cite{Sharkey2017} or by \cite{Wadsworth2010} improves inference significantly.
Still, \cor{our} orthogonal parameterization is even more efficient, especially for the estimation of the tail parameter $\xi$: 
the autocorrelation reduces even more rapidly with the lag, and the ESS increases faster with the number of draws.
With the recommendations of $\text{ESS} \geq 400$ for estimation \citep{gelman2013bayesian}, our experimental setup is satisfactory only in the orthogonal case because of $\xi$. 
In contrast, more iterations are required to fulfill this condition for the parameterization recommended by  \cite{Sharkey2017}.

\begin{figure}
    \includegraphics[trim={0cm 0cm 0cm 0cm},clip,width =1\textwidth]{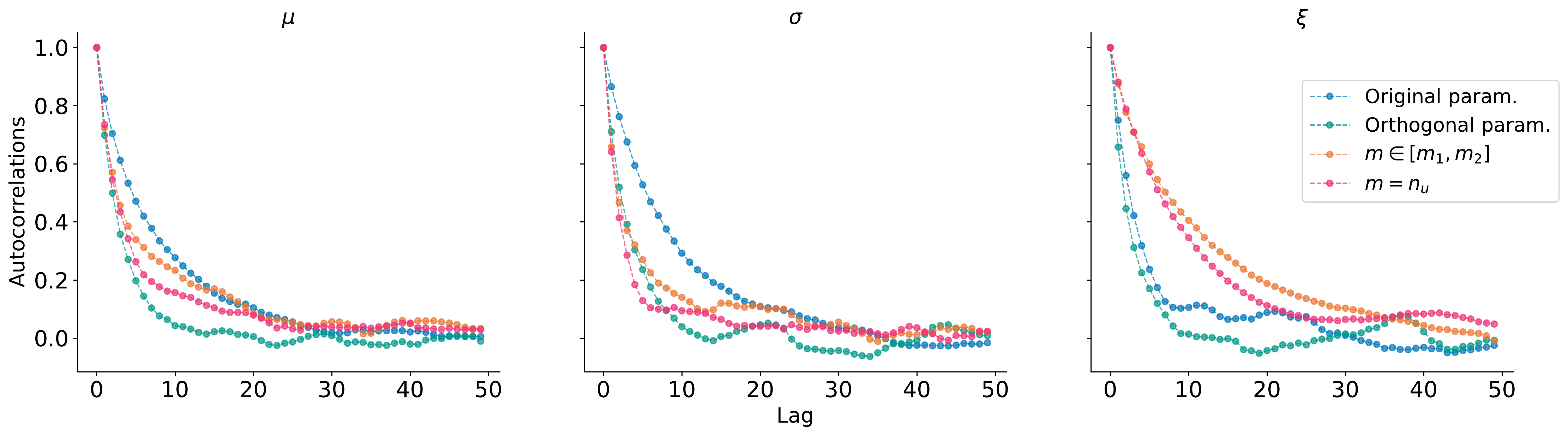}

    \includegraphics[trim={0cm 0cm 0cm 0cm},clip,width =1\textwidth]{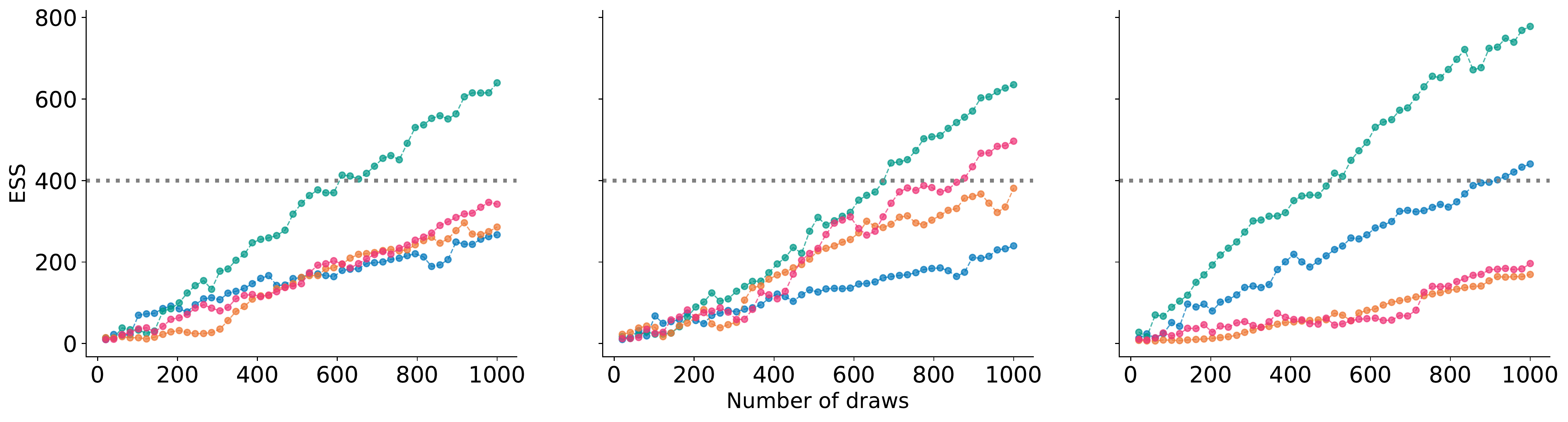}

    \includegraphics[trim={0cm 0cm 0cm 0cm},clip,width =1\textwidth]{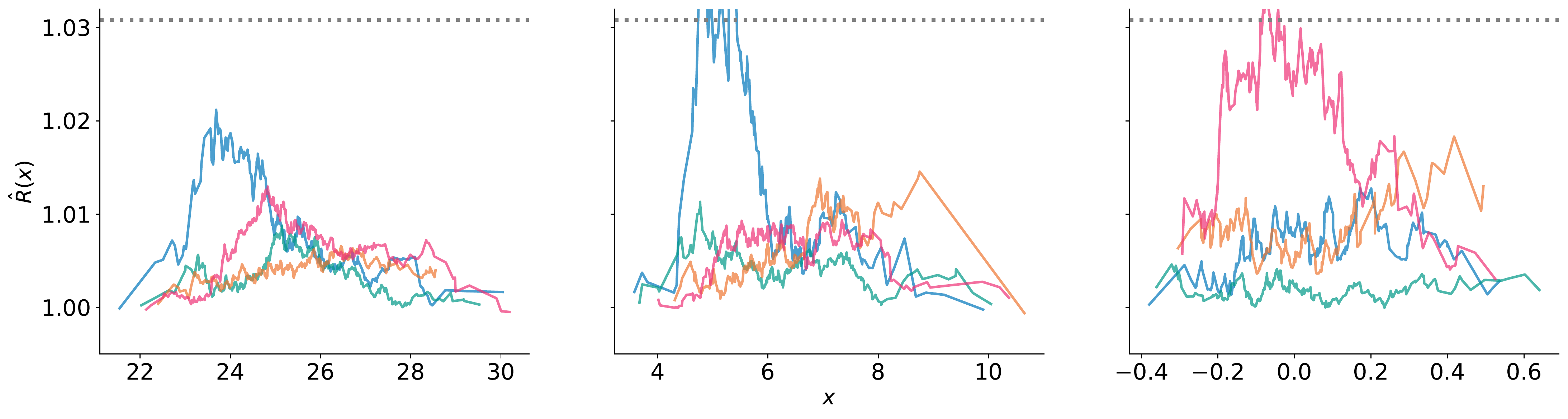}
    \caption{Convergence diagnostic plots for Poisson parameters $(\mu, \sigma, \xi)$ with $\xi = 0$, after $1\,000$ Metropolis--Hastings draws and a burn-in of $1\,000$, for four different parameterizations: the original one (in \textcolor{colorblind1}{blue}), the \cite{Sharkey2017} update with $m \in [\hat{m}_1, \hat{m}_2]$ (in \textcolor{colorblind3}{orange}), the \cite{Wadsworth2010} update with $m = n_u$ (in \textcolor{colorblind4}{magenta}), and the orthogonal parameterization (in \textcolor{colorblind2}{green}).
    Top row: autocorrelations as functions of the lag. 
    Second row: evolution of ESS with the number of draws (the gray line corresponds to value of $400$ recommended in \cite{gelman2013bayesian}).
    Bottom row: $\hat{R}(x)$ as a function of the quantile $x$, with the adapted threshold of $1.03$ \citep[see][]{moinsRhatArxiv}.}
    \label{fig:PP_xi_nul}
\end{figure}

\paragraph{Example with $\xi = 0$}
Finally when $\xi = 0$, the GPD and therefore the intensity $\Lambda(I_u)$ of the Poisson process defined in Section~\ref{subsec:bayes_ext_intro} reduce to an exponential model with location and scale parameters.
Figure~\ref{fig:PP_xi_nul} shows an example in this case with 
$(m, u, \mu, \sigma, \xi) = (20, 20, 25, 5, 0)$, leading to $r \approx 54$ expected observations.
Similarly to the case $\xi > 0$ in Section~\ref{subsec:simulations}, this example illustrates that updating $m$ like \cite{Sharkey2017} or \cite{Wadsworth2010} is beneficial for MCMC convergence, but less than using orthogonal parameterization.
In the same way as in the two other maximum domains of attraction, this parameterization is the most efficient one for the convergence of Metropolis--Hastings algorithm. 

\subsection{GPD and GEV case}
\label{subsec:gev_gpd}

\begin{figure}
    \centering
    \includegraphics[trim={0cm 0cm 0cm 0cm},clip,width =0.7\textwidth]{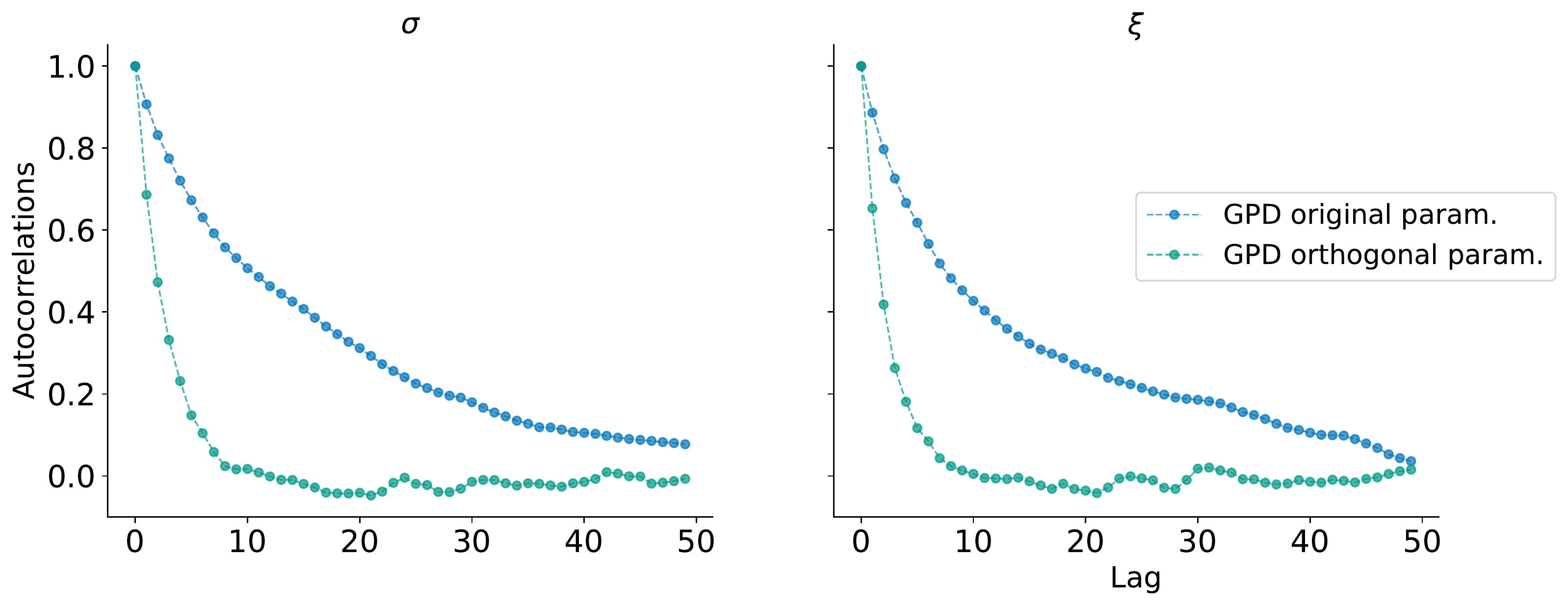}

    \includegraphics[trim={0cm 0cm 0cm 0cm},clip,width =0.7\textwidth]{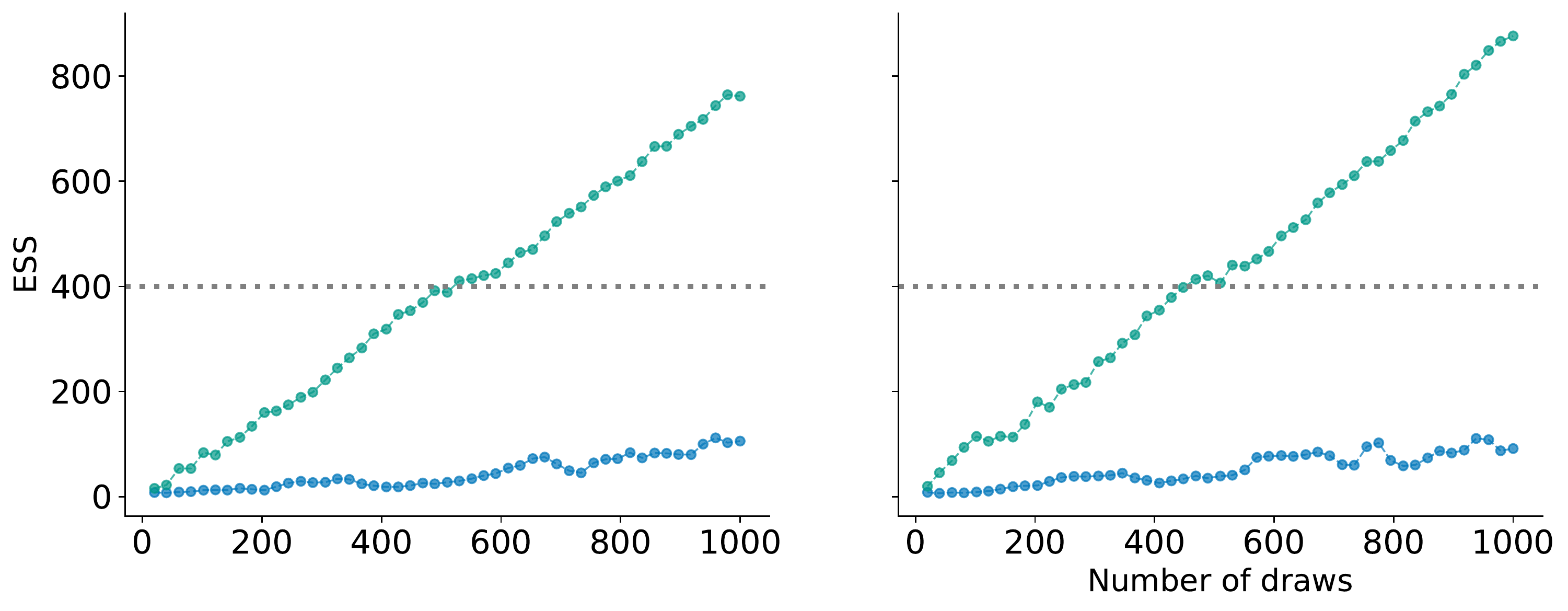}

    \includegraphics[trim={0cm 0cm 0cm 0cm},clip,width =0.7\textwidth]{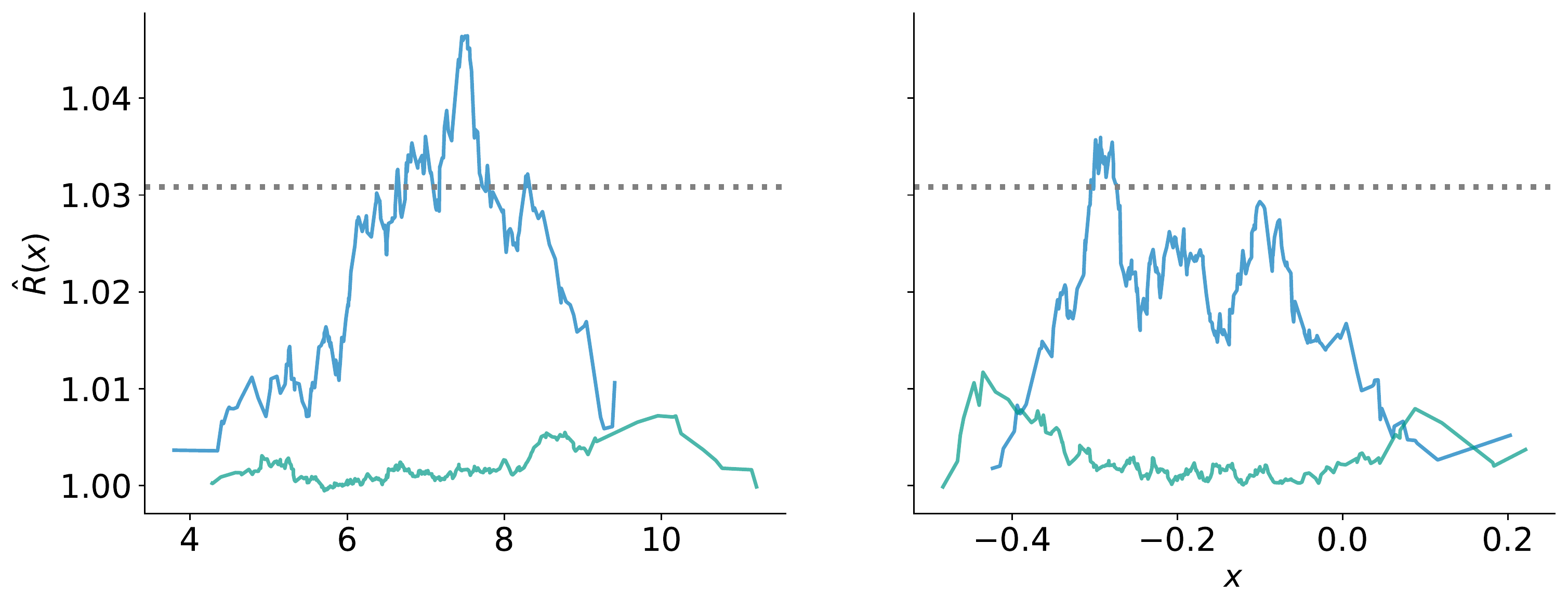}
    \caption{Convergence diagnostic plots for GPD parameters $(\sigma, \xi)$ with $\xi < 0$, after $1\,000$ Metropolis--Hastings draws and a burn-in of $1\,000$, for two parameterizations, the original (in \textcolor{colorblind1}{blue}) and the orthogonal one (in \textcolor{colorblind2}{green}).
    Top row: autocorrelations as functions of the lag. 
    Second row: evolution of ESS with the number of draws (the gray line corresponds to value of $400$ recommended in \cite{gelman2013bayesian}).
    Bottom row: $\hat{R}(x)$ as a function of the quantile $x$, with the adapted threshold of $1.03$ \citep[see][]{moinsRhatArxiv}.}
    \label{fig:GPD_xi_neg}
\end{figure}

In the particular case of GPD (defined in Equation~(\ref{eq:gpd_distribution})) that arises in the traditional peaks over threshold model, the same observation can be made about the benefits of an orthogonal parameterization for $(\sigma, \xi)$.
More precisely, the transformation $(\nu, \xi) = \left(\sigma(1+ \xi), \xi\right)$ leads to an orthogonal Fisher information matrix for GPD \citep{Chavez-Demoulin2005}, and improves MCMC convergence as shown in Figure~\ref{fig:GPD_xi_neg}.
The same experimental setup as in the Poisson process case is used here, with a choice of $(\sigma, \xi) = (5, -0.1)$ and $u = 25$.
Again, all plots in Figure~\ref{fig:GPD_xi_neg} show that the chains mixing is satisfactory only in the case of an orthogonal parameterization, while the original parameterization requires more iterations to be effective for inference.
Up to our knowledge, there is no orthogonal parameterization for the GEV likelihood known in the literature.
However, it should be noted that the parameters of the Poisson process model $(\mu, \sigma, \xi)$ correspond to those of the block maxima framework with $m$ blocks (see Section~\ref{subsec:bayes_ext_intro}).
Consequently, we should expect a similar convergence issue for parameters $(\mu, \sigma, \xi)$ with GEV likelihood, and therefore an improvement in the MCMC convergence with the use of the orthogonal parameterization $(r, \nu, \xi)$ of the Poisson model.

\subsection{Ratio-of-uniforms}
\label{subsec:ration_uniform}
\begin{figure}
    \centering
    \small\textcolor{colorblind1}{\textbf{Original parameters} $(\mu,\sigma,\xi)$} ($p_a \approx 0.195$)\\ \vspace{-0.4cm}
    \includegraphics[trim={0cm 0cm 0cm 0cm},clip,width =1\textwidth]{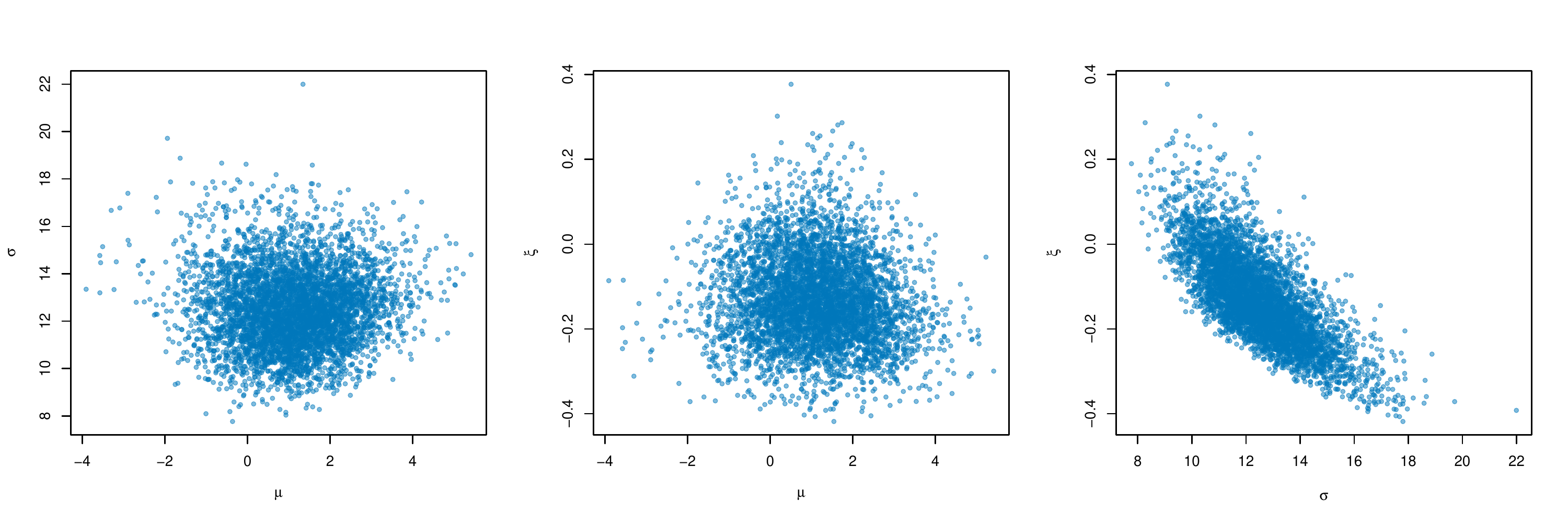}

    \textcolor{colorblind4}{\textbf{Original parameters  $(\mu,\sigma,\xi)$ with $m=n_u$}} ($p_a \approx 0.284$)\\ \vspace{-0.4cm}
    \includegraphics[trim={0cm 0cm 0cm 0cm},clip,width =1\textwidth]{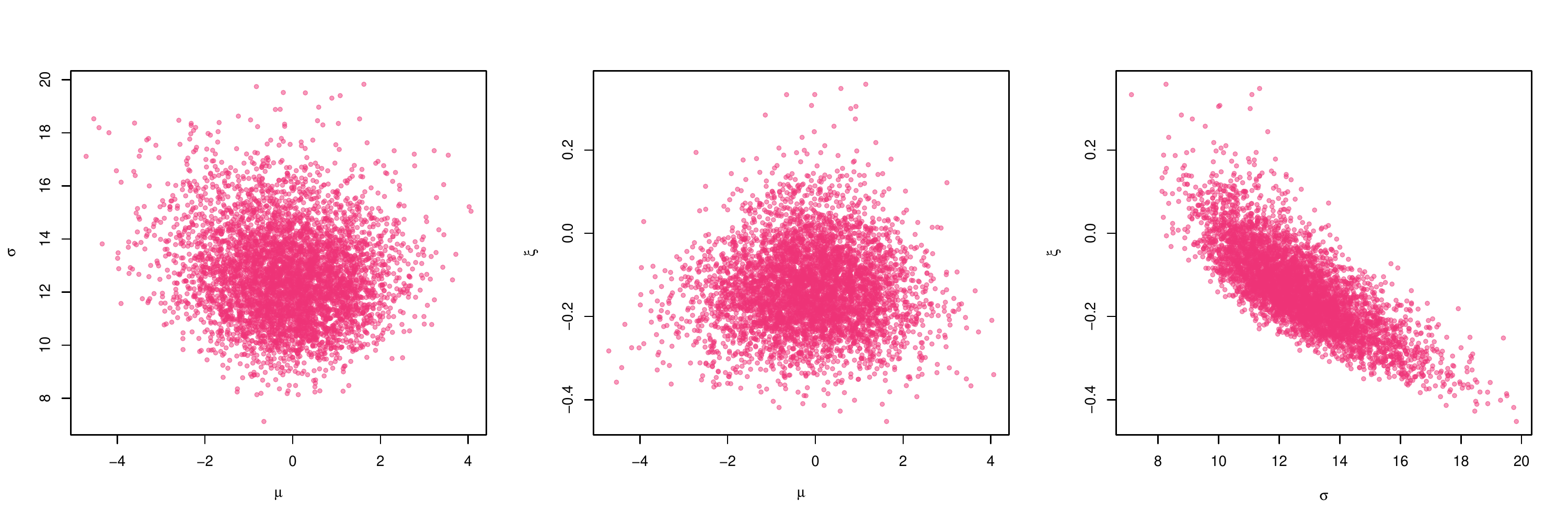}

    \textcolor{colorblind2}{\textbf{Orthogonal parameters} $(r,\nu,\xi)$} ($p_a \approx 0.301$)\\ \vspace{-0.4cm}
    \includegraphics[trim={0cm 0cm 0cm 0cm},clip,width =1\textwidth]{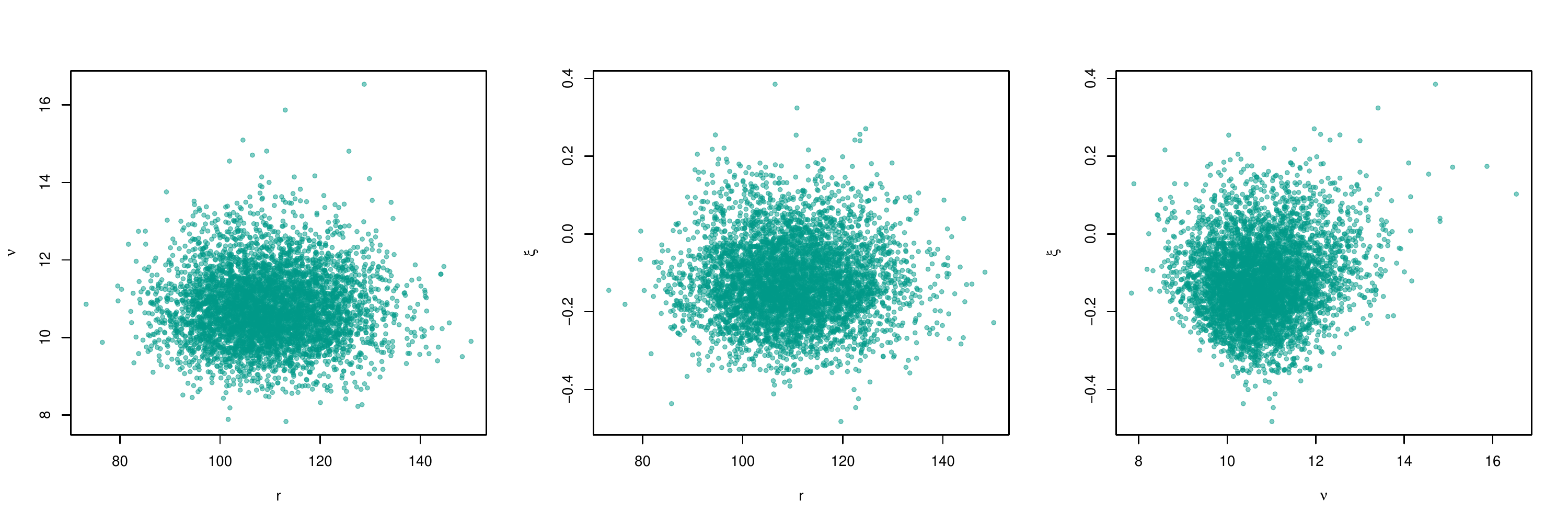}
    \caption{Pairwise plots of parameter values simulated using the ratio of uniform method, for three parameterizations:
    the original one (in \textcolor{colorblind1}{blue}), the \cite{Wadsworth2010} update with $m = n_u$ (in \textcolor{colorblind4}{magenta}), and the orthogonal parameterization (in \textcolor{colorblind2}{green}).
    The probability of acceptance $p_a$ reflects the efficiency of the sampling method.
    }
    \label{fig:ratio_neg}
\end{figure}

\cor{
The benefits of reparameterization for Bayesian inference can be extended to other sampling methods  than MCMC.
Typically, the efficiency of acceptance-rejection algorithms can be altered if the geometry of the acceptance region is too complex, and this can be due to correlation between parameters.
The \texttt{rust} package \citep{rust} implements such an acceptance-rejection algorithm dedicated to extreme value models, the so-called generalized ratio-of-uniforms method. 
It consists in simulating uniformly values in a region that encloses an acceptance region, where the ratio of the obtained samples is distributed according to the target distribution (see \citet[][Chapter~5]{gilks1995markov} for more details).
As explained in the \texttt{revdbayes} documentation \citep{revdbayes} which is built upon \texttt{rust}, the efficiency of this method highly depends on the probability of acceptance $p_a$.
\texttt{revdbayes} already includes the possibility to use the reparameterization suggested by \cite{Wadsworth2010} with $m=n_u$ for the Poisson process, along with a rotate option to reduce dependence.
We add the orthogonal parameterization in the comparison, and show the results in Figure~\ref{fig:ratio_neg}.
We set $(m, u, \mu, \sigma, \xi) = (100, 0, 1, 1, -0.1)$, and draw $10\,000$ samples for three configurations.
As expected, the orthogonal parameterization slightly improves the probability of acceptance compared to the case where $m=n_u=110$, which is already significantly better than the case where $m$ is not changed ($m=100$).
Note that this package operates a transformation of variable before sampling to improve normality, which in view of the bottom row in Figure~\ref{fig:ratio_neg}, may not be necessary for the orthogonal parameters.}

\subsection{Replications and comparison with maximum likelihood}
\label{sec:replications_mse}

Despite the \cor{fact that the} Bayesian paradigm comes with several benefits (briefly described in Section~\ref{subsec:bayes_ext_intro}), one can be interested in the comparison with frequentist estimator \cor{such as} maximum likelihood estimation (MLE).
From a frequentist point of view, this involves extracting a pointwise estimator from the posterior distribution, \cor{such as} the posterior mean, and replicate the experiment to estimate the mean squared error (MSE).
The two steps of the Bayesian workflow we study here are expected to impact the performance of these estimators.
A parameterization which leads to poor convergence of the MCMC chains will affect the accuracy of estimation, and the prior can add a bias that may or may not be advantageous to the estimation.

\begin{figure}
    \centering
    \begin{tabular}{cc}
        % Estimations with $n=100$ samples & Estimations with $n=500$ samples\\ 
        \includegraphics[trim={0cm 0cm 0cm 0cm},clip,width =0.45\textwidth]{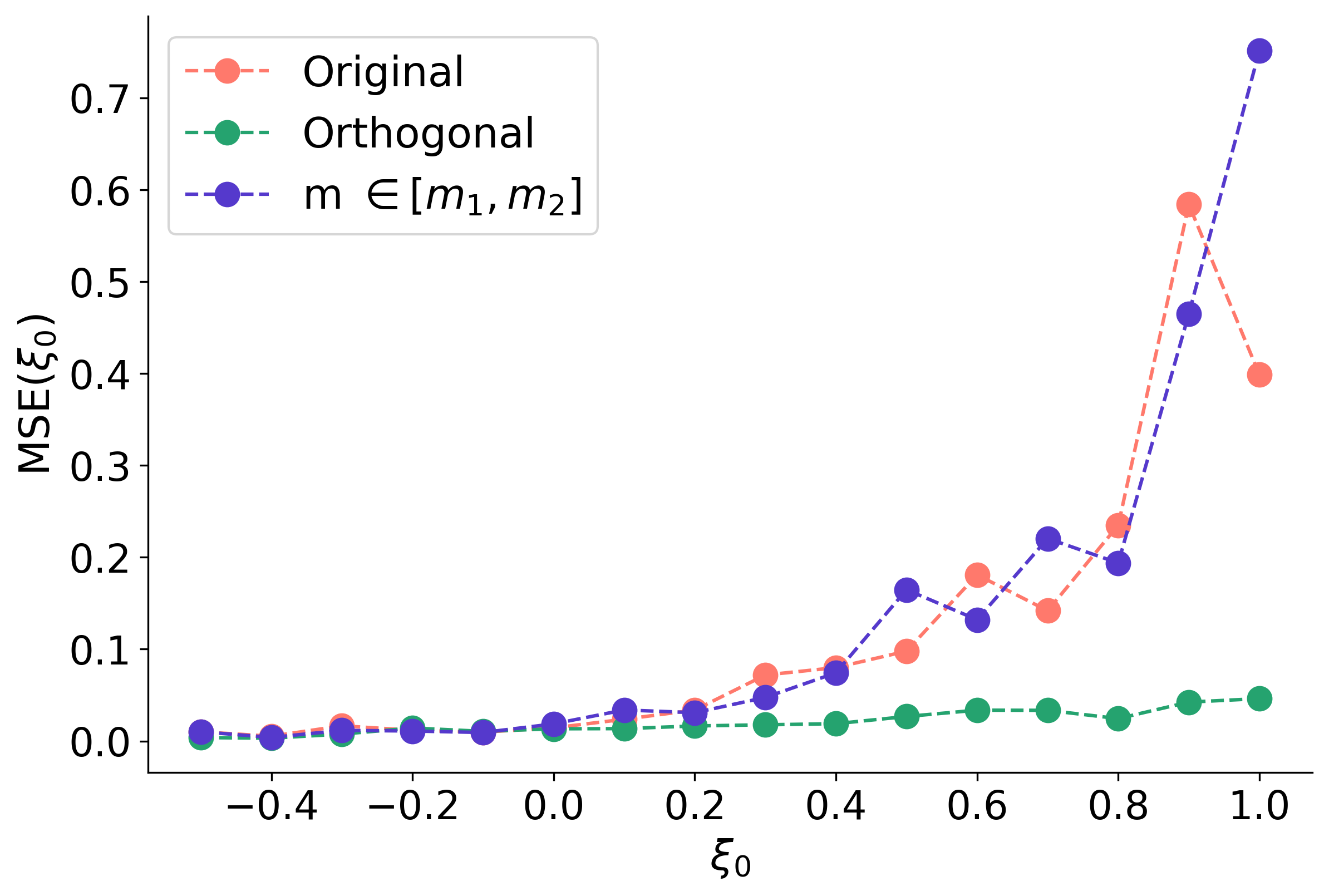}
        &
        \includegraphics[trim={0cm 0cm 0cm 0cm},clip,width =0.45\textwidth]{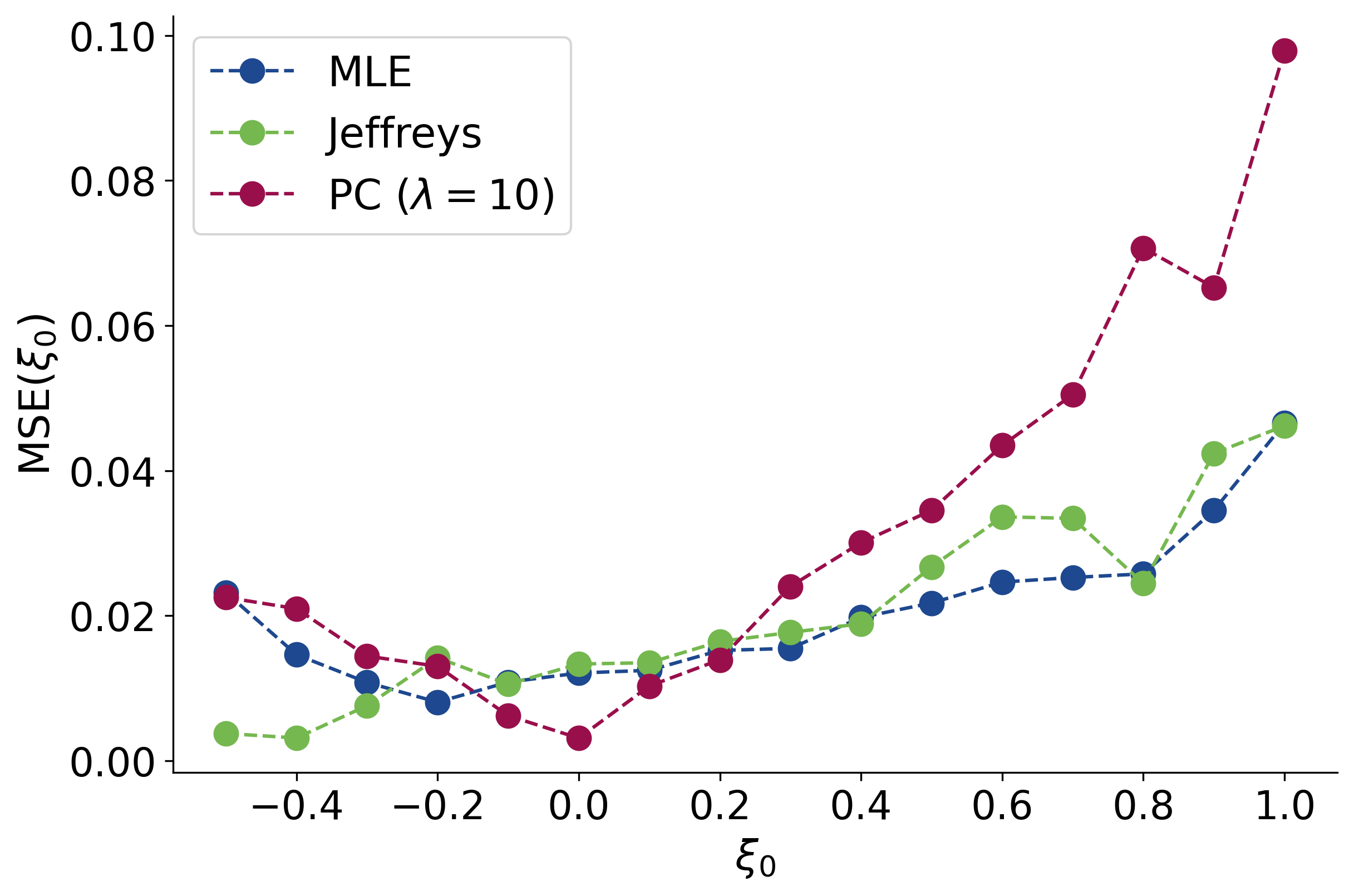}
    \end{tabular}
    \caption{Mean squared error (MSE) on the estimation of $\xi$ for a true value $\xi_0 \in [-1/2, 1]$. 
    The computation is done on $100$ replications for each value of $\xi_0$.
    Left panel: different parameterizations under Jeffreys prior.
    Right panel: \cor{Jeffreys and PC priors under orthogonal parameterization, along with MLE}.
    }
    \label{fig:mse}
\end{figure}

For different values of $\xi_0$ between $-0.5$ and $1$, we replicate $100$ times the following experiment (this range includes a large number of models and allows \cor{us} to have both Jeffreys and PC priors well defined): for $i = 1, \ldots, 100$,
we generate samples $\boldsymbol{x}_i$ according to a Poisson process distribution with parameters $(m, u, \sigma, \xi) = (1, 10, 15, \xi_0)$ and $\mu$ in a way such that the expected number of points is equal to $r=100$: 
\begin{equation*}
    \mu = u - \frac{\sigma}{\xi_0} (100^{-\xi_0}-1).
\end{equation*}
Then, we run MCMC chains with the same configuration as in Section~\ref{sec:exp} and compute the posterior mean $\hat{\xi}_i = \mathbb{E}[\xi \mid \boldsymbol{x}_i]$.
We these $100$ experiments, we compute the MSE:
\begin{equation*}
    \text{MSE}(\xi_0) = \frac{1}{100} \sum_{i=1}^{100} (\hat{\xi}_i - \xi_0)^2.
\end{equation*}
First, we compare the different parameterizations for the Poisson process with the same Jeffreys prior. 
Results are displayed in the left panel of Figure~\ref{fig:mse}, and illustrate the inaccuracy of the frameworks without reparameterization and with the update of \cite{Sharkey2017}, due to lack of convergence of MCMC. 
This issue is getting worse as $\xi_0$ increases, and a bias/variance decomposition of the MSE shows that it is mostly due to the variance term.
Then, for the same orthogonal parameterization, we compare Jeffreys prior, PC prior with a choice of $\lambda = 10$, and the MLE for the Poisson process, implemented using the extRemes package \citep{gilleland2016extremes}. 
Results in the right panel of Figure~\ref{fig:mse} show that the performance of the posterior mean estimation with Jeffreys prior is approximately the same as the MLE, except when $\xi_0$ is \cor{close to} $-1/2$ where the asymptote behaviour of Jeffreys favours the estimation. 
This shows that, despite the uninformative construction, this prior can favour negative values of $\xi_0$ close to $-1/2$.
The behaviour of PC prior is, as expected, penalizing the values of $\xi$ far from $\xi_0 = 0$.
When $\xi_0$ is around zero, this prior outperforms Jeffreys' one and MLE, but assuming a value near zero when $|\xi_0|$ is large can add a large bias.

% \end{appendices}

\end{document}